\documentclass[twocolumn,aps,prl,superscriptaddress,floatfix]{revtex4-2}  
\usepackage{graphicx}  
\usepackage{dcolumn}   
\usepackage{bm}        
\usepackage{amssymb}   
\usepackage{amsmath}   
\usepackage[version=4]{mhchem}
\usepackage{siunitx}
\usepackage{natbib}
\usepackage{braket}
\usepackage{color} 

\begin{document}

\title{Emergence of a spin microemulsion in spin-orbit coupled Bose-Einstein condensates}
\author{Ethan C. McGarrigle} 
\email{emcgarrigle@ucsb.edu}
\affiliation{Department of Chemical Engineering, University of California, Santa Barbara, California 93106, USA}

\author{Kris T. Delaney} 
\affiliation{Materials Research Laboratory, University of California, Santa Barbara, California 93106, USA}

\author{Leon Balents}
 \affiliation{Kavli Institute for Theoretical Physics, University of California, Santa Barbara, California 93106, USA}
\affiliation{Canadian Institute for Advanced Research, Toronto, Ontario, Canada}
 
\author{Glenn H. Fredrickson}
\email{ghf@ucsb.edu}
\affiliation{Department of Chemical Engineering, University of California, Santa Barbara, California 93106, USA}
\affiliation{Materials Research Laboratory, University of California, Santa Barbara, California 93106, USA}
\affiliation{Materials Department, University of California, Santa Barbara, California 93106, USA}


\begin{abstract}
We report the first numerical prediction of a ``spin microemulsion''\textemdash a phase with undulating spin domains resembling classical bicontinuous oil-water-surfactant emulsions \textemdash in two-dimensional systems of spinor Bose-Einstein condensates with isotropic Rashba spin-orbit coupling. Using field-theoretic numerical simulations, we investigated the melting of a low-temperature stripe phase with supersolid character and find that the stripes lose their superfluidity at elevated temperature and undergo a Kosterlitz--Thouless-like transition into a spin microemulsion. Momentum distribution calculations highlight a thermally broadened occupation of the Rashba circle of low-energy states with macroscopic and isotropic occupation around the ring. We provide a finite-temperature phase diagram that positions the emulsion as an intermediate, structured isotropic phase with residual quantum character before transitioning at higher temperature into a structureless normal fluid. 
\end{abstract}

\maketitle

\textit{Introduction.}\textemdash Microemulsion phases are typically found in synthetic soft matter systems \cite{lee_bulk_1995, guo_all-aqueous_2023,  bonelli_lamellar_2019}, most prominently in oil-water-surfactant mixtures \cite{huang_bicontinuous_2017, de_gennes_microemulsions_1982, clausse_bicontinuous_1981}. Two immiscible species form enriched domains while often a third minority component stabilizes their interfaces, playing the role of a surfactant or amphiphile. Another example is ternary polymer blends, where two incompatible homopolymers can form a bicontinuous microemulsion in the presence of a high-molecular-weight diblock copolymer \cite{bates_polymeric_1997, spencer_coexistence_2021}. Beyond the classical soft matter context, microemulsion-like phases are proposed as intermediates between Wigner crystals and Fermi liquids \cite{spivak_transport_2006, spivak_phases_2004} in 2D electronic systems such as metal-oxide-semiconductor field-effect transistors \cite{radzihovsky_quantum_2009, berg_charge-4e_2009, sun_fluctuating_2008}. 

A microemulsion analogue phase is plausible for bosons with isotropic two-dimensional (2D) ``Rashba'' spin-orbit coupling (SOC). A continuous circle with radius $q_0$ of degenerate energy minima appears in the single-particle dispersion and constitutes a massive ground-state degeneracy, analogous to 2D Rashba electronic systems \cite{bychkov_oscillatory_1984, michiardi_optical_2022, wu_two-dimensional_2021}. The circular degeneracy provides a setting where isotropic density modulations with dominant length scale $2 \pi / q_0$ can produce an isotropic, emulsion-like phase. 

Although most experimental realizations of SOC in cold atom systems access anisotropic or one-dimensional SOC \cite{lin_spinorbit-coupled_2011, khamehchi_spin-momentum_2016, ji_experimental_2014, hamner_spin-orbit-coupled_2015, li_stripe_2017, bersano_experimental_2019}, a recent experiment from the NIST group \cite{valdes-curiel_topological_2021} achieved a Rashba-like 2D SOC, but did not quite achieve a circular single-particle degeneracy. Exotic behavior has been speculated: that Rashba bosons could host a Bose metal or Rashba Luttinger liquid \cite{sur_metallic_2019}, exhibit fractional Hall-like states \cite{valdes-curiel_topological_2021}, or undergo a statistical transmutation into composite fermions \cite{sedrakyan_composite_2012}. Bogoluibov theory, mean-field theory, and semi-classical numerical simulations \cite{liao_searching_2018, kawasaki_finite-temperature_2017, sinha_trapped_2011, jian_paired_2011, ozawa_condensation_2013, stanescu_spin-orbit_2008} provide mounting evidence that bulk systems of Rashba bosons will produce a plane-wave or stripe ground state near $T=0$, where Bose condensation occurs into one or two minima at non-zero momenta, respectively. However, their finite temperature physics has eluded conclusive theoretical understanding due to a sign problem inherent to the SOC Hamiltonian.  

In this Letter, we interrogate a second-quantized model of interacting, pseudospin-1/2 Rashba bosons at finite temperature using an approximation-free, field-theoretic simulation technique that overcomes the sign problem. Our investigation reveals a novel microemulsion phase with undulating pseudospin domains reflecting an isotropic occupation of the Rashba circle of low-energy states. We perform simulations at several SOC strengths and temperatures to compute a phase diagram and confirm the microemulsion's existence between a low-temperature superfluid spin-stripe phase and a high-temperature paramagnetic, normal fluid.

\textit{Methods.}\textemdash We incorporate 2D isotropic SOC into a model of interacting pseudospin-1/2 bosons via a uniform and time-independent non-Abelian gauge potential $\bm{\mathcal{A}} = \hbar \kappa (\underline{\underline{\sigma}}^{x} , \underline{\underline{\sigma}}^{y} ) $ \cite{goldman_light-induced_2014}, where $\underline{\underline{\sigma}}^{\nu}$ are the spin-1/2 Pauli-matrices in the $\nu$ direction and $\kappa$ is the SOC strength. Assemblies of interacting pseudospin-1/2 bosons in 2D are well described by the second-quantized Hamiltonian: 
\begin{equation} 
  \begin{split} 
   & \hat{H} = \sum_{\alpha \gamma} \int d^2 r \hspace{2px} \hat{\psi}^{\dagger}_{\alpha} (\mathbf{r}) \left [ \frac{1}{2m} (\mathbf{\hat p} \underline{\underline{I}} - \bm{\mathcal{A} })^2  - \mu \underline{\underline{I}} \right]^{\vphantom{\dagger}}_{\alpha \gamma} \hspace{-3px} \hat{\psi}^{\vphantom{\dagger}}_{\gamma} (\mathbf{r})  \\
    &+ \frac{1}{2} \sum_{\alpha \gamma} \int d^2 r \hspace{2px} \hat{\psi}^{\dagger}_{\alpha} (\mathbf{r}) \hat{\psi}^{\dagger}_{\gamma} (\mathbf{r}) (g_0 \underline{\underline{I}} + g_1 \underline{\underline{\sigma}}^{x} )_{\alpha \gamma} \hat{\psi}^{\vphantom{\dagger}}_{\alpha} (\mathbf{r}) \hat{\psi}^{\vphantom{\dagger}}_{\gamma} (\mathbf{r}),
     \label{eq: H} 
     \end{split}
\end{equation} 
 \noindent{where} $\hat{\psi}_{\alpha} (\mathbf{r})$ is a second-quantized field operator for pseudospin species $\alpha$ obeying Bose commutation relations \cite{fetter_quantum_2012}, $\mathbf{\hat p} = -i \hbar \nabla$ is the canonical momentum operator, $\underline{\underline{I}}$ is the 2x2 identity matrix for pseudospin indices, $\mu$ is chemical potential, and $m$ is the atomic mass shared by both pseudospin species. We assume contact pairwise interactions where the repulsive coupling constants $g_0$ and $g_1$ are parameterized by s-wave scattering lengths $a_{s, 0}$ and $a_{s, 1}$ for like and unlike pseudospin scattering events, respectively, and correspond to a symmetric matrix of coupling constants. 
 
We define a natural length scale $\ell = \sqrt{\frac{\hbar^2}{2m \mu_{\text{eff} } } }$ and energy scale $\mu_{\text{eff}}$, where $\mu_{\text{eff}}$ is an effective chemical potential that incorporates the SOC recoil energy $\mu_{\text{eff}} \equiv \mu - \hbar^2 \kappa^2 / m$. After rescaling all field operators, spatial derivatives, and integrals using $\ell$ and $\mu_{\text{eff}}$, we identify four dimensionless quantities that govern the model's universal behavior: a repulsion scale $\tilde{g} \equiv 2m g_0 / \hbar^2 $, a SOC strength $\tilde{\kappa} \equiv \kappa \ell $, a temperature $\tilde{T} \equiv k_{B} T / \mu_{\text{eff}} $, and a miscibility parameter $\eta_{g} \equiv  g_{1} / g_{0} $. For this study, we confine our simulations to slightly immiscible conditions $\eta_{g} = 1.1 > 1$, which promote a stripe ground state at low temperatures. 

To access finite temperature, we use Feynman's imaginary time path integral approach to convert the many-boson problem into a coherent state quantum field theory in the grand canonical ensemble \cite{negele_quantum_1988}. The degrees of freedom are a complex conjugate pair of bosonic coherent state fields $\phi_{\alpha} (\mathbf{r}, \tau)$, $\phi^*_{\alpha} (\mathbf{r}, \tau)$ for each pseudospin species, obeying periodic boundary conditions in imaginary time $\tau$. The resulting grand partition function $ \mathcal{Z} = \int \mathcal{D} (\bm{\phi^{\vphantom{*}}}, \bm{ \phi^*} ) \hspace{2px} e^{-S [\bm{\phi^{\vphantom{*}}}, \bm{\phi^*} ]} $ consists of functional integrals over the real and imaginary parts of the field components at each space-imaginary time point; $\bm{\phi^{\vphantom{*}}}$ and $\bm{\phi^*}$ are 2-component vectors in the pseudospin species. A configuration's statistical weight is described by $e^{-S}$, where $S[\bm{\phi^{\vphantom{*}}}, \bm{\phi^*}]$ is a complex-valued action functional that presents a numerically foreboding sign problem: 
\begin{widetext}
  \begin{align}
   S[\bm{\phi}, \bm{\phi}^*] &= 
 \sum_{\alpha} \sum_{j=0}^{N_{\tau} -1} \int d^2 r \hspace{3px}  \phi^*_{\alpha, j} (\mathbf{r}) [ \phi^{\vphantom{*}}_{\alpha, j} (\mathbf{r}) -  \phi^{\vphantom{*}}_{\alpha, j-1} (\mathbf{r}) ] + \frac{\tilde{\beta}}{N_{\tau}}  \sum_{\alpha, \gamma} \sum_{j=0}^{N_{\tau} -1} \int d^2 r \hspace{3px} \phi^*_{\alpha, j} (\mathbf{r}) \hspace{1.5px}  \mathcal{\hat K}_{\alpha, \gamma} \hspace{1.5px} \phi^{\vphantom{*}}_{\gamma, j-1} (\mathbf{r}) \\
   &+ \frac{\tilde{\beta} \tilde{g} }{2 N_{\tau}}   \sum_{\alpha, \gamma} \sum_{j=0}^{N_{\tau} -1} \int d^2 r \hspace{3px} \phi^*_{\alpha, j} (\mathbf{r}) \phi^*_{\gamma, j} (\mathbf{r}) (\underline{\underline{I}} + \eta_{g} \underline{\underline{ \sigma}}^{x} )^{\vphantom{*}}_{\alpha \gamma} \phi^{\vphantom{*}}_{\gamma, j-1} (\mathbf{r}) \phi^{\vphantom{*}}_{\alpha, j-1} (\mathbf{r}), \nonumber
       \label{eq: S}
     \end{align} 
    \end{widetext} 
\noindent{where} $\tilde{\beta} = 1 / \tilde{T}$, imaginary time $\tau \in [0, \tilde{\beta}]$ has been discretized into $N_{\tau}$ points, and the index $j$ labels the imaginary-time slice. We enforce periodic boundary conditions in imaginary time, e.g. $\phi_{\alpha, 0} (\mathbf{r}) = \phi_{\alpha, N_{\tau}} (\mathbf{r})$, as well as in the two spatial dimensions. $\hat{\underline{\underline{\mathcal{K}}}}$ is a Hermitian one-body matrix with equal diagonal entries $\hat{\mathcal{K}}_{\uparrow \uparrow} = \hat{\mathcal{K}}_{\downarrow \downarrow} = -\tilde{\nabla}^2 -1 $, and $\hat{\mathcal{K}}_{\uparrow \downarrow} = \hat{\mathcal{K}}^{\dagger}_{\downarrow \uparrow} = -2 \tilde{\kappa} [-i \partial_{\tilde{x}} - \partial_{\tilde{y}}]$   , where $\partial_{\tilde{x}}$ and $\tilde{\nabla}^2$ indicate a dimensionless spatial derivative and Laplacian, respectively. $\mathbf{r}$ is a dimensionless coordinate. 

In our field-theoretic simulation method, we discretize the coherent state fields in the three space-imaginary time dimensions and numerically sample the field elements by evolving the complex Langevin algorithm \cite{delaney_numerical_2020} 
\begin{equation}
   \begin{split}
  \frac{\partial}{\partial t}  \phi^{\vphantom{*}}_{\alpha, j} (\mathbf{r}, t) &= - \frac{\delta S [\phi^{\vphantom{*}}, \phi^*]}{\delta \phi^*_{\alpha, j} (\mathbf{r}, t) }  + \eta^{\vphantom{*}}_{\alpha, j} (\mathbf{r}, t) \\
  \frac{\partial}{\partial t} \phi^*_{\alpha, j} (\mathbf{r}, t) &= - \frac{\delta S  [\phi^{\vphantom{*}}, \phi^*]}{\delta \phi^{\vphantom{*}}_{\alpha, j} (\mathbf{r}, t) } + \eta^*_{\alpha, j} (\mathbf{r}, t) ,
  \label{eq: CL}
    \end{split}
\end{equation}
forward in fictitious time $t$ using an exponential time-differencing integrator \cite{fredrickson_field-theoretic_2023}. $\eta^{\vphantom{*}}_{\alpha, j} = \eta^{(1)}_{\alpha, j} + i \eta^{(2)}_{\alpha, j}$ and $\eta^*_{\alpha, j} = \eta^{(1)}_{\alpha, j} - i \eta^{(2)}_{\alpha, j}$ are complex conjugate white noise sources that are built from real noise fields $\eta^{(1)}_{\alpha, j}$ and $\eta^{(2)}_{\alpha, j}$ with $\langle \eta^{(\mu)}_{\alpha, j} (\mathbf{r}, t) \rangle = 0 $ and $\langle \eta^{(\mu)}_{\alpha, j} (\mathbf{r}, t) \eta^{(\nu)}_{\beta, k } (\mathbf{r'}, t') \rangle = \delta_{\mu, \nu} \delta_{\alpha, \beta} \delta_{j, k}  \delta (\mathbf{r} - \mathbf{r'}) \delta (t - t') $. Our pseudospectral implementation assumes periodic boundary conditions in all dimensions. Observables of interest such as the pressure and internal energy are obtained by averaging field operators (functionals of the coherent state fields) over fictitious complex-Langevin time trajectories. This procedure overcomes the statistical weight's highly oscillatory nature and provides bias-free thermodynamic properties at finite temperature \cite{parisi_complex_1983, klauder_langevin_1983}. 

To identify the equilibrium phase, we calculated snapshots of the resulting structure's real-space density profile as well as thermal averages (via fictitious Langevin time averages) of the momentum state distribution in $\mathbf{k}$-space, where $\mathbf{k} = 2 \pi (n_{x} / L_{x} , n_{y} / L_{y} ) $ is the discrete wavevector labelling momentum states with $n \in \mathbb{Z}$. The density profile for pseudospin species $\alpha$ and the momentum distribution were calculated via $\rho_{\alpha} [\bm{\phi^{\vphantom{*}}} , \bm{\phi^*} ; \mathbf{r}] = 1 / N_{\tau}  \sum_{j=0}^{N_{\tau}-1} \phi^{*}_{\alpha, j} (\mathbf{r})  \phi_{\alpha, j-1} ^{\vphantom{*}} (\mathbf{r}) $ and $N[\bm{\phi^{\vphantom{*}}}, \bm{\phi^*} ; \mathbf{k}] = A/N_{\tau} \sum_{\alpha} \sum_{j=0}^{N_{\tau}-1} \tilde{\phi}^*_{\alpha, j, \mathbf{-k}}  \tilde{\phi}^{\vphantom{*}}_{\alpha, j-1, \mathbf{k}} $, where $\tilde{\phi}$ refers to the Discrete Fourier Transform of $\phi$, accessed numerically via Fast Fourier Transform algorithms \cite{delaney_numerical_2020}, and $A = L_{x} L_{y}$ is the system size (area). 

 \begin{figure}[ht] 
\includegraphics[scale = 0.40]{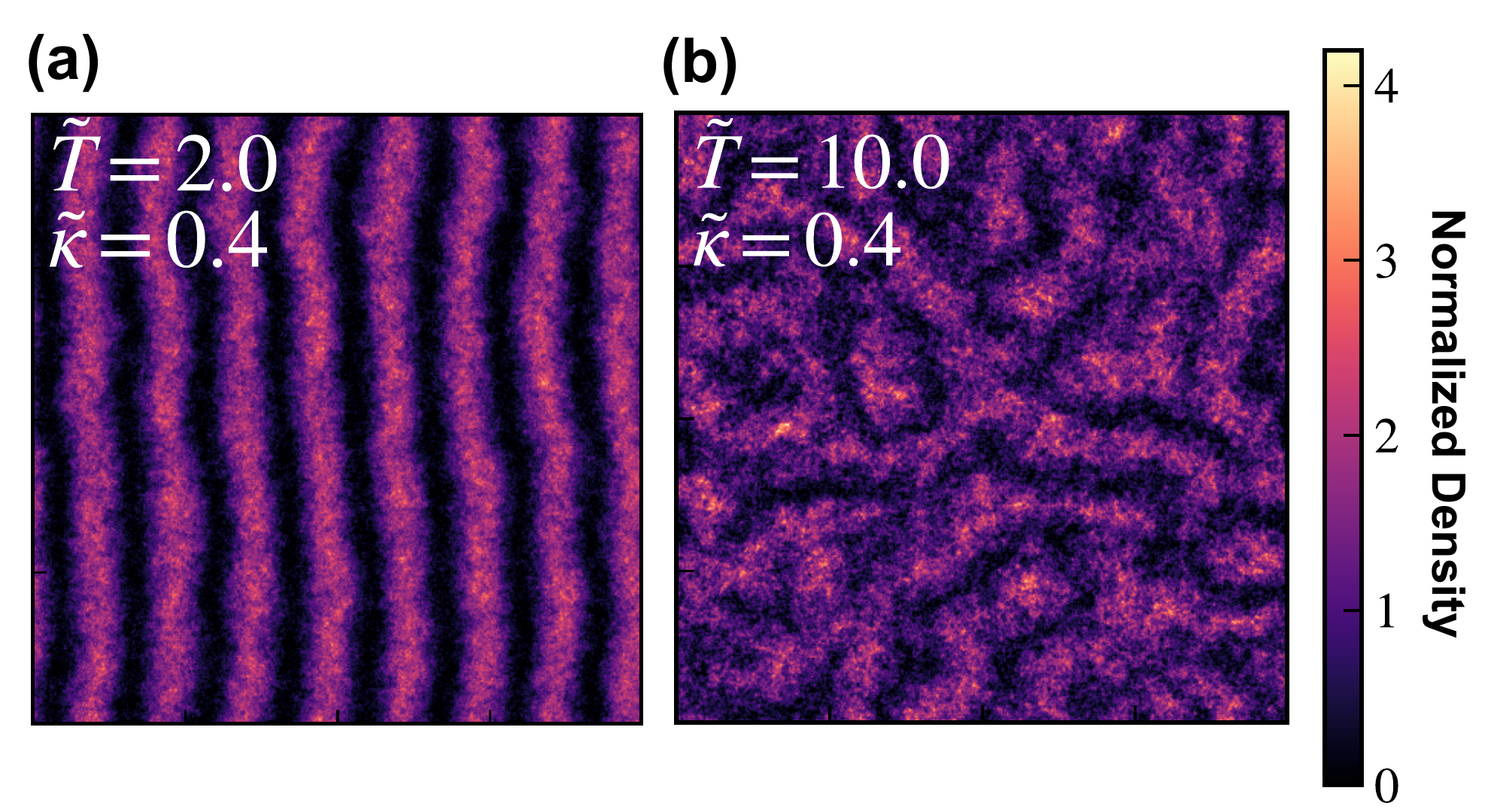} \label{fig: 1a} 
 \caption{Density profiles $\rho_{\uparrow} (\mathbf{r})$ of pseudospin bosons in the $\ket{\uparrow}$ state at $\tilde{\kappa} = 0.4$ and interaction conditions $\tilde{g} = 0.05$, and $\eta_{g} = 1.1$. (a) - (b) The stripe phase with Rashba SOC transitions into an isotropic spin microemulsion above $\tilde{T}_{c}$. Profiles are snapshots normalized by the spatially averaged density.}  \label{fig: 1}
 \end{figure}

\begin{figure}[ht]
    \includegraphics[scale = 0.42]{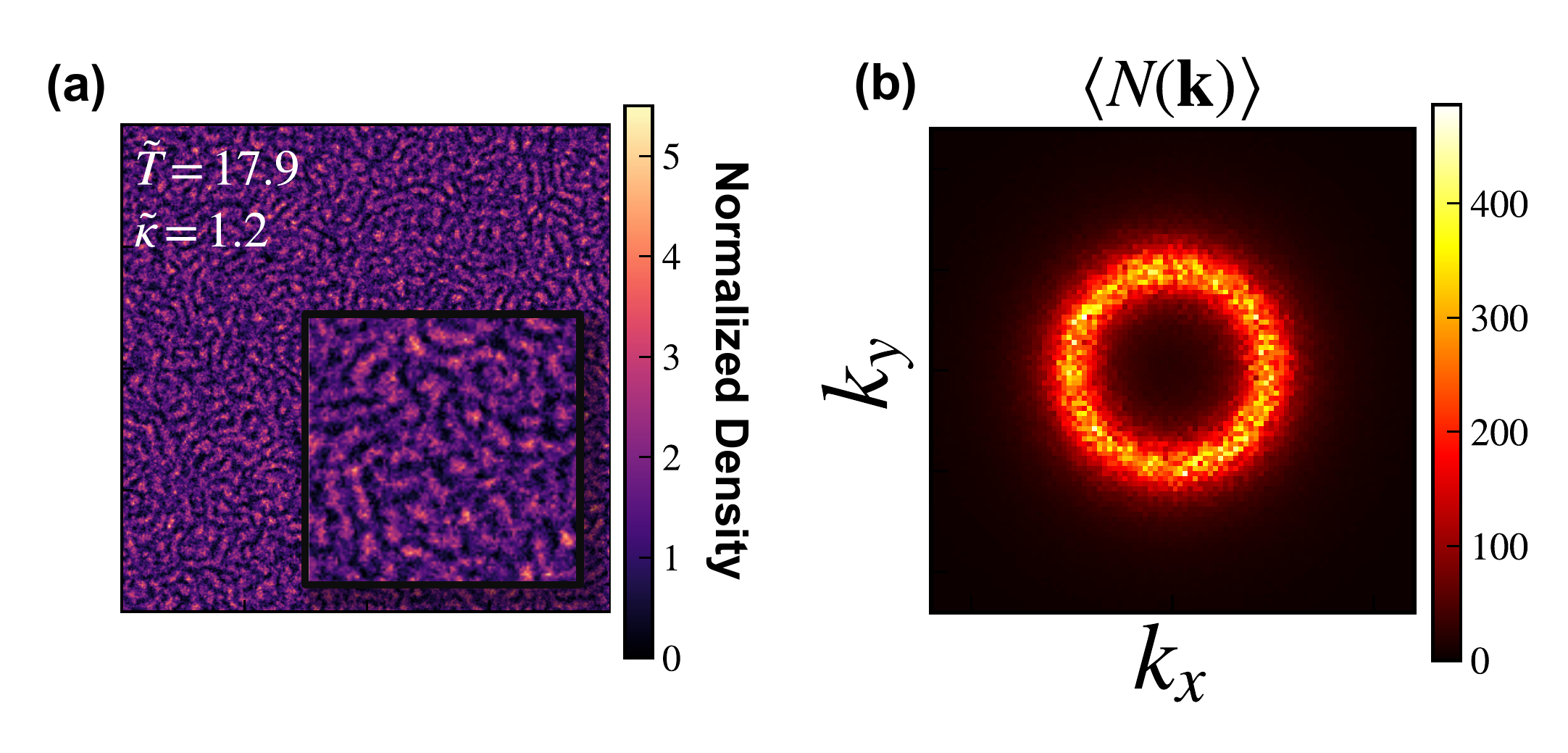}
 \caption{A spin microemulsion phase with $\langle N \rangle \sim \num{4.8e5}$ particles at $\tilde{T} = 17.9$, $\tilde{\kappa} = 1.2$, $\tilde{g} = 0.05$, and $\eta_{g} = 1.1$. a)  Density profile $\rho_{\uparrow} (\mathbf{r})$ of pseudospin bosons in the $\ket{\uparrow}$ basis state, normalized by the spatially averaged density. A highlighted section is enlarged to show the local emulsion-like structure. b) Thermal average momentum state occupation $N(\mathbf{k})$.} 
  \label{fig: 2}
  \end{figure}

\textit{Appearance of a microemulsion.}\textemdash At low temperatures and immiscible conditions ($\eta_{g} > 1$), the system possesses a stripe ground state with broken continuous translational and rotational symmetries, shown in Figure\ (\ref{fig: 1}a). The stripe phase is a pseudospin density wave in one direction that resembles spin density waves found in electronic systems \cite{sun_fluctuating_2008}. However, in the presence of isotropic SOC and above a critical temperature, the spin stripes undulate and twist into an isotropic ``spin emulsion'' structure that resembles bicontinuous microemulsions found in soft matter systems (Figure\ (\ref{fig: 1}b)). Figure\ (\ref{fig: 1}) qualitatively shows this transition from a quasi-long range ordered stripe phase to an isotropic spin microemulsion via $\ket{\uparrow}$ density snapshots at different temperatures. Figure\ (\ref{fig: 2}a) provides a wide field of view of a larger microemulsion system with no quasi-long range orientational order. 

The momentum distribution is shown in Figure\ (\ref{fig: 2}b) and highlights the isotropic character of the structured emulsion phase with characteristic domain width $\pi / 2 \tilde{\kappa}$. The significant occupation of momentum states with wavevector $|\mathbf{k}| = \tilde{\kappa}$ in Figure\ (\ref{fig: 2}b) suggests a massive fragmentation of the quasi-condensate onto the Rashba dispersion's circular manifold of states. This circular momentum distribution presents an appealing experimental signature for the microemulsion phase and could be observed in a time-of-flight cold atom experiment, using either spin-resolved (Stern-Gerlach) or spin-unresolved techniques. 

\begin{figure}[ht]
 \includegraphics[scale = 0.38]{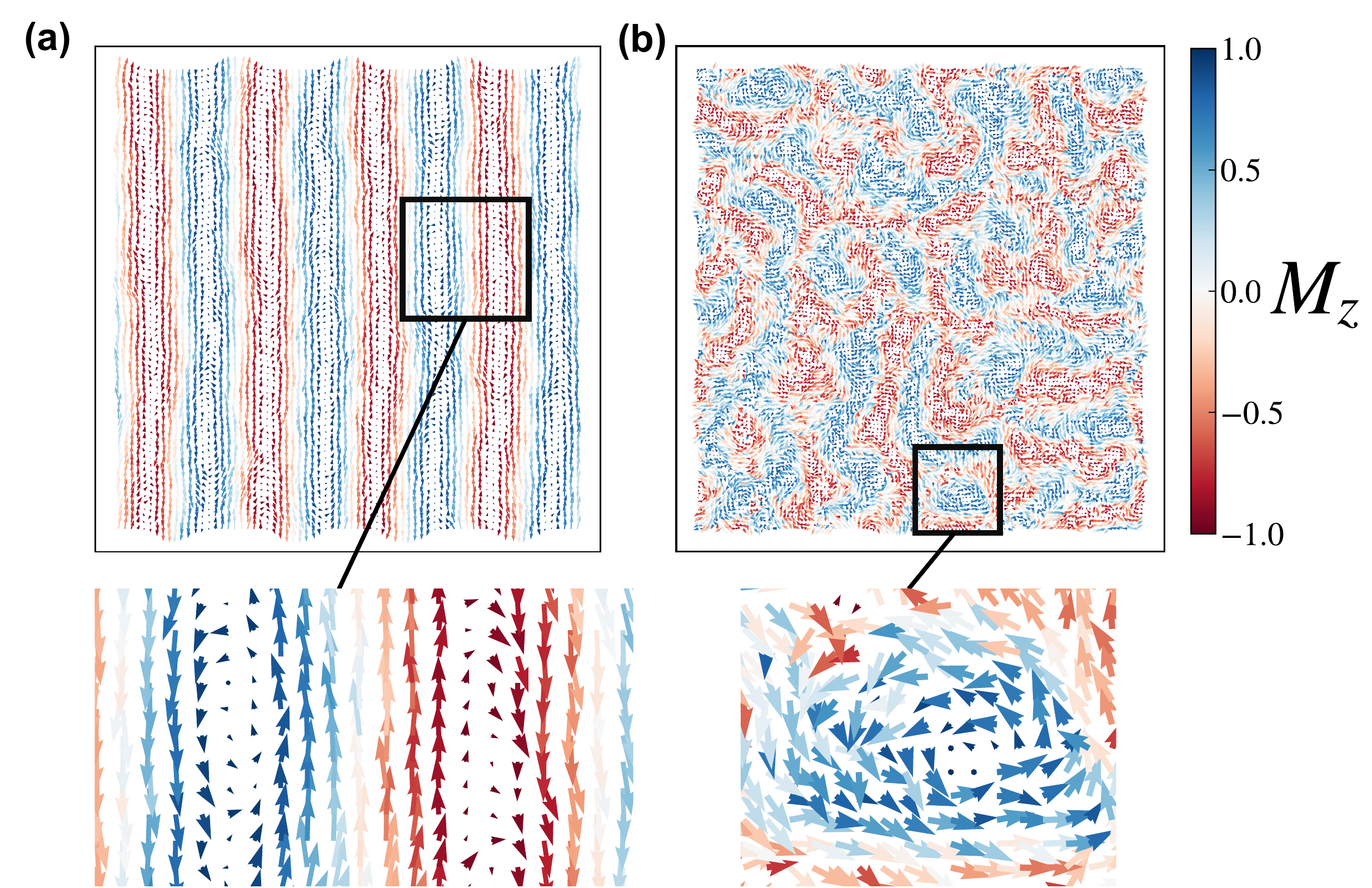} 
  \caption{Pseudospin textures and topological defects visualized via $\mathbf{M} (\mathbf{r})$ at $\tilde{\kappa} = 0.4$, $\tilde{g} = 0.05$, and $\eta_{g} = 1.1$ in (a) the stripe phase, $\tilde{T} = 1.1$, and (b) the microemulsion phase, $\tilde{T} = 10.25$. Color bar shows the magnitude of $M_{z} (\mathbf{r})$. Arrows represent the planar vector $(M_{x} (\mathbf{r}), M_{y} (\mathbf{r}) )$. Boxed regions are enlarged to show planar pseudospin domain walls and vortices for the stripe and microemulsion phases, respectively. The plotted local magnetization vector is normalized by the local magnitude $| M (\mathbf{r}) | = \sqrt{M_{x}^2 (\mathbf{r}) + M_{y}^2 (\mathbf{r}) + M_{z}^2 (\mathbf{r}) }$.} 
   \label{fig: 3}
\end{figure} 

To study the local pseudospin details, we calculate components of the local magnetization field via $M_{\nu} [\bm{\phi} , \bm{\phi^*} ; \mathbf{r}] = \frac{1}{N_{\tau}} \sum_{\alpha \beta} \sum_{j = 0}^{N_{\tau} - 1}  \phi^*_{\alpha, j} (\mathbf{r}) \sigma^{\nu}_{\alpha \beta} \phi^{\vphantom{*}}_{\beta, j-1} (\mathbf{r})$, 
where $\sigma^{\nu}_{\alpha \beta}$ represents an element of the spin-1/2 Pauli matrix in the $\nu$ direction. Both the stripe and microemulsion phases are characterized by strong $M_{z} (\mathbf{r})$ domains shown in the color intensity of Figure\ (\ref{fig: 3}). Furthermore, the stripe and emulsion phase exhibit rich $M_{x} (\mathbf{r})$ and $M_{y} (\mathbf{r})$ planar pseudospin textures. The zoomed regions in Figure\ (\ref{fig: 3}a) and (\ref{fig: 3}b) highlight planar pseudospin defects present throughout both phases. The stripe phase hosts stable domain walls and 1D skyrmions commensurate with the stripe periodicity; however, thermal fluctuations induce pairs of oppositely charged planar pseudospin vortices between adjacent stripes (Figure\ (\ref{fig: 3}a)); in contrast, vortices in the microemulsion phase appear free and manifest as 2D skyrmions (Figure\ (\ref{fig: 3}b)). The shift from bound to free vortices across the stripe to microemulsion transition suggests a Kosterlitz--Thouless picture. These vortices in pseudospin disappear completely upon crossing over at higher temperatures to the paramagnetic normal fluid. 

\begin{figure}[ht] 
\includegraphics[scale = 0.50]{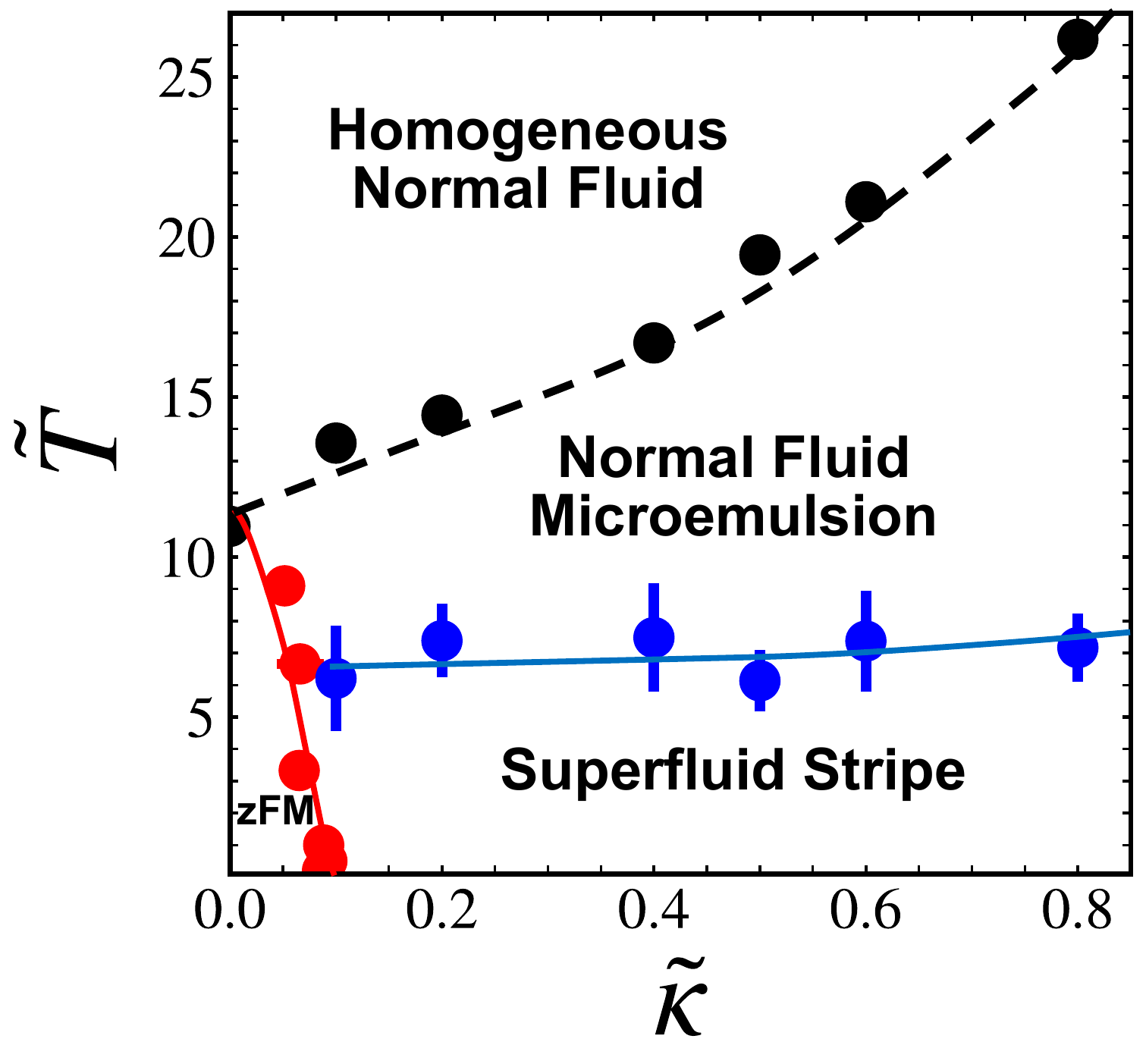}
 \caption{Finite-temperature phase diagram in the $\tilde{T}$--$\tilde{\kappa}$ plane of interacting bosons with isotropic SOC in 2 dimensions, constructed for $\eta_{g} = 1.1$, $\tilde{g} = 0.05$. Lines are a guide for the eye, and error bars are discussed in the Supplementary Material \cite{SM_ref} (see also references \cite{mukerjee_topological_2006, kobayashi_berezinskii-kosterlitz-thouless_2019, radic_stoner_2014} therein). The blue solid line denotes the Kosterlitz--Thouless-like critical transition, while the red line denotes a first order phase transition into the $\hat{z}$-Ferromagnet (zFM) phase. The black dashed line depicts a crossover between the spin microemulsion and homogeneous normal fluid. }  \label{fig: 4}
 \end{figure}

\begin{figure}[ht]
\includegraphics[scale = 0.35]{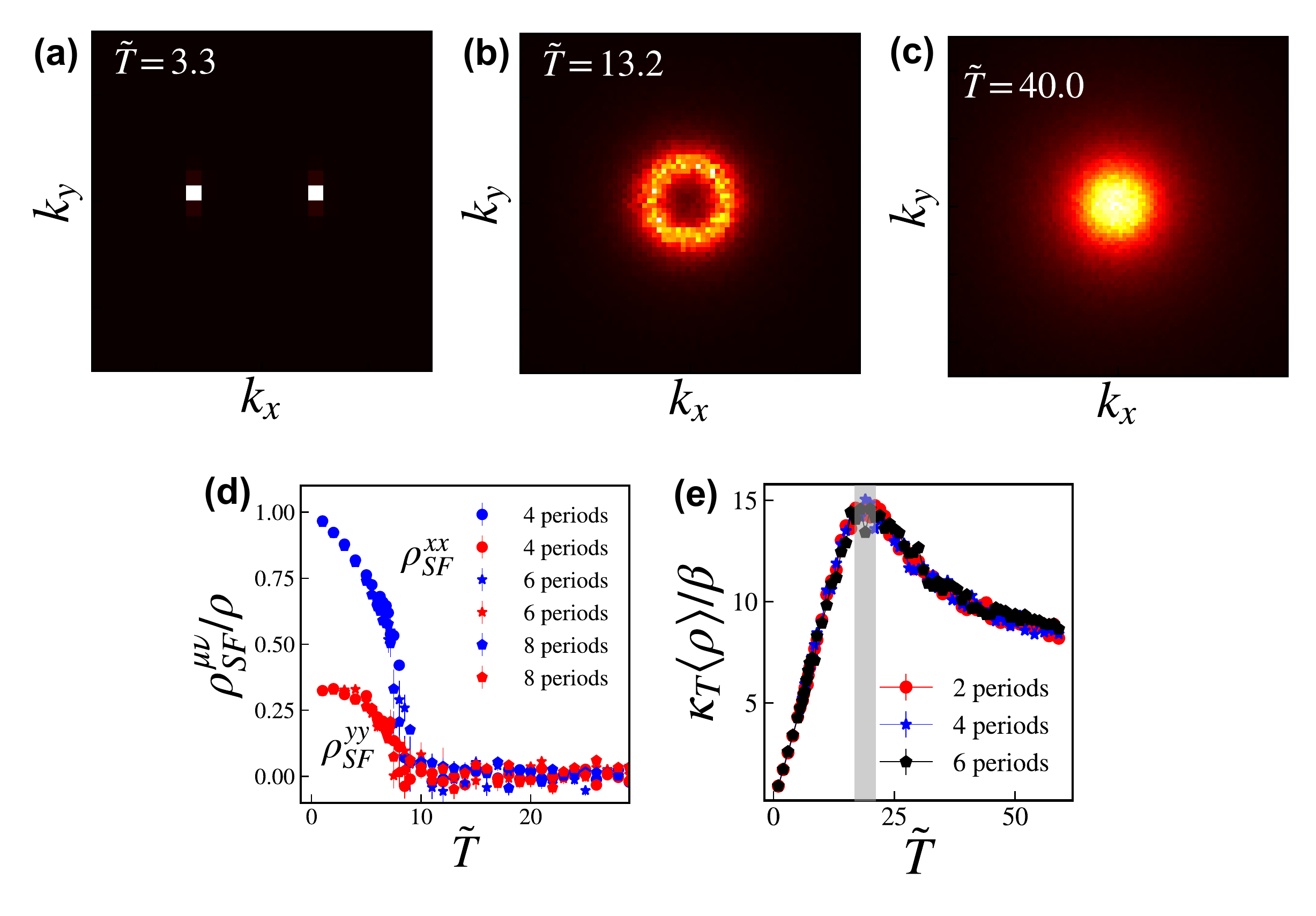} 
  \caption{Finite-temperature phase transitions at $\tilde{\kappa} = 0.5$, $\tilde{g} = 0.05$, and $\eta_{g} = 1.1$. (a) - (c) Momentum distribution $N(\mathbf{k})$ for the (a) stripe phase at $\tilde{T} = 3.3$, (b) spin microemulsion phase at $\tilde{T} = 13.2$, and (c) homogeneous normal fluid at $\tilde{T} = 40.0$. (d) Superfluid stiffness tensor across the stripe to emulsion thermal phase transition at various system sizes. The $\rho^{xx}_{\text{SF}}$ ($\rho^{yy}_{\text{SF}}$) fraction is shown with blue (red) markers. (e) Isothermal compressibility at different system sizes, scaled for data collapse. The universal peak near $\tilde{T} = 19.4$ (gray shading) highlights the crossover from a structured microemulsion to a homogeneous normal fluid. All error bars are standard errors of the mean calculated during the Langevin time sample averaging process.}    
    \label{fig: 5}
\end{figure}

\textit{Phase Transitions}.\textemdash The predicted finite-temperature phase diagram in the $\tilde{T}$--$\tilde{\kappa}$ plane is shown in Figure\ (\ref{fig: 4}) for slightly immiscible and isotropic SOC conditions. To observe various thermal transitions, we varied the temperature at fixed $\tilde{\kappa} \geq 0.1$ and monitored the thermal-averaged momentum distributions, which revealed three distinct phases, shown in Figures\ (\ref{fig: 5}a -- \ref{fig: 5}c). Different interaction strengths $\tilde{g} > 0$ and miscibility values $\eta_{g} > 1$ would shift phase boundaries while leaving the topology of the phase diagram unaltered. The microemulsion emerges as a clear and robust intermediate between the low-temperature ordered stripe superfluid and a high-temperature homogeneous normal fluid.

Remarkably, the microemulsion's appearance at moderate $\tilde{T}$ coincides with a complete loss of superfluidity (Figure\ (\ref{fig: 5}d)) from the stripe phase, showing characteristics similar to a Kosterlitz-Thouless transition. To determine that phase boundary for a given $\tilde{\kappa}$, the superfluid stiffness tensor's diagonal normal component $ \rho^{x x}_{\text{SF}}$ was tracked with increasing temperature for $\hat{x}$-modulated stripe phases. The superfluid density tensor is estimated using the phase twist method \cite{subasi_quantum-geometric_2022} $\rho^{\mu \nu}_{\text{SF}} = \frac{m}{\hbar^2 A} \frac{\partial^2 \Omega (\mathbf{q})}{\partial q_{\mu} \partial q_{\nu}}|_{\mathbf{q} \to 0 }$, where $\Omega (\mathbf{q}) $ is the thermodynamic grand potential after an imposed super-flow proportional to $\mathbf{q}$, incorporated via a pseudospin and $\tau$-independent phase shift to each coherent state field via $\phi_{\alpha} (\mathbf{r}, \tau) \to \phi_{\alpha} (\mathbf{r}, \tau) e^{i \mathbf{q} \cdot \mathbf{r}} $. The field operator functional used to calculate $\rho^{\mu \nu}_{\text{SF}}$ is detailed in the Supplementary Material \cite{SM_ref}.  Using our dimensionless framework and superfluid density calculations, we estimate the stripe-microemulsion transition temperature in $^{87}\ce{Rb}$ condensates to be near 150 nK, assuming a mixture of $\ket{F=1, m_{F} = \pm 1}$ hyperfine states with an effective chemical potential of $\mu_{\text{eff}} = h \times 0.95$ \text{kHz} \cite{vankempenInterisotopeDeterminationUltracold2002}.

For vertically oriented stripes under isotropic SOC, the stripe phase is nearly pure superfluid in $\hat{x}$ but significantly normal fluid  in $\hat{y}$; free particle-like excitations appear in the $\hat{y}$ Bogoluibov spectrum \cite{liao_searching_2018} and disrupt superfluidity along the stripes. In contrast, the $\hat{x}$ excitation spectrum \cite{liao_searching_2018} hosts double gapless bands that signify supersolid character normal to the stripes and arise from the simultaneously broken $U(1)$ and continuous translational symmetries. 

$\rho^{x x}_{\text{SF}}$ declines to zero near the critical stripe melting temperature (Figure\ (\ref{fig: 5}d)), reminiscent of the finite-size Kosterlitz-Thouless transition in conventional 2D superfluids, liquid crystals, and superconductors \cite{nelson_universal_1977}. To determine the precise critical temperature, we conducted a finite-size scaling analysis in the Supplemental Material \cite{SM_ref} on $ \rho^{x x}_{\text{SF}} / \rho $ to correct for finite-size errors and estimate the Kosterlitz-Thouless transition temperature $\tilde{T}_{c}$ in the thermodynamic limit. The normal component $ \rho^{x x}_{\text{SF}} / \rho$ fraction experiences a full variation and serves as a more appropriate measure than $\rho_{\text{SF}} / \rho$ of the helicity or superfluidity modulus for quantifying a universal jump. Despite the stripe's apparent smectic character observed in a modest simulation cell, the finite-temperature stripe phase in the thermodynamic limit should lose long range positional order in the presence of quantum and thermal fluctuations to produce a 2D superfluid nematic \cite{jian_paired_2011}, analogous to the phonon and defect-mediated smectic to nematic transition in classical 2D liquid crystals \cite{toner_smectic_1981, hammond_temperature_2005}. Therefore, we expect that the observed transition would be a Kosterlitz-Thouless transition between a superfluid nematic and the normal fluid microemulsion in the thermodynamic limit. Our finite-size analysis suggests a transition mediated by the unbinding of half-vortex phase defects in the nematic superfluid, where the magnitude of the universal jump is modified \cite{jian_paired_2011}. 

At higher temperatures, the spin microemulsion continuously crosses over to a disordered paramagnetic state. On the single-particle level, this crossover overcomes an energy difference $\Delta \sim   \tilde{\kappa}^2$ between the upper helicity branch and the degenerate circle on the lower helicity branch of the Rashba dispersion; this scaling supports the positive curvature of the microemulsion to homogeneous fluid crossover curve observed (Figure\ (\ref{fig: 4})). The homogeneous fluid represents a significant occupation of the single-particle branch with positive helicity and is supported by the momentum distribution in Figures\ (\ref{fig: 5}b) and (\ref{fig: 5}c), where the maximum occupation moves from $\mathbf{k} = |\tilde{\kappa}|$ to $\mathbf{k} = \mathbf{0}$ as $\tilde{T}$ increases. This crossover coincides with an anomaly in the 2D isothermal compressibility $\kappa_{T} = -\frac{1}{A} \frac{dA}{dP}|_{T}$ (Figure\ (\ref{fig: 5}e)) that shows no statistically significant finite-size error, suggesting a crossover rather than a critical phase transition with diverging correlation length. The field operator functional used to calculate $\kappa_{T}$ is detailed in the Supplementary Material.

At low SOC strength $\tilde{\kappa} < 0.1$ and low $\tilde{T}$, the system reduces to a weakly interacting pseudospin-1/2 Bose gas, where the immiscibility promotes a $\hat{z}$-Ferromagnet (zFM) superfluid ground state with broken $\mathbb{Z}_2$ and $U(1)$ symmetries. This state is conveniently probed in the grand canonical ensemble where the system selects a nearly full occupation of either pseudospin state $\ket{\uparrow}$ or $\ket{\downarrow}$ due to a thermodynamic phase coexistence of two homogeneous superfluids with opposing macroscopic pseudospin. The transition between the zFM and the stripe phase was studied via direct calculation of the grand potential $\Omega$ for each phase \cite{fredrickson_direct_2022}, where the phase with the lowest grand potential is deemed more thermodynamically stable. The Supplementary Material \cite{SM_ref} includes observed jumps in order parameters such as the $\mathbf{k}=\mathbf{0}$ occupation fraction $N_{k=0} / N$ and the integrated $\hat{z}$-projection of the magnetization $M_{z} = \int d^2 r\hspace{2px} M_{z} (\mathbf{r})$, which suggest that the phase transition is first order and is confirmed by an observed kink in the grand potential at $\tilde{\kappa}_{c}$.
 
\textit{Conclusions.}\textemdash We have discovered an isotropic spin microemulsion phase, found in Rashba spin-orbit coupled, Bose-Einstein condensates at finite temperature. Our Letter details its density profile, pseudospin profile, and momentum distribution, computed using complex Langevin sampling of a representative coherent state field theory. The microemulsion exhibits short-range correlations in z-pseudospin magnetization on a length scale $\ell \sim \pi / \tilde{\kappa} $, consistent with the Rashba circle of degenerate single-particle states at momentum $|\mathbf{k}| = \tilde{\kappa}$. This circular momentum distribution is reminiscent of other intermediate isotropic phases in condensed matter physics, such as the melting of Dzyaloshinskii--Moriya three-dimensional helimagnets \cite{noauthor_phys_nodate-1} and itinerant magnets \cite{zhang_infinite_2023}. We quantify the superfluid fraction and find that the microemulsion is entirely normal fluid; however, the robust spinor domains highlight the microemulsion's quantum character despite losing quasi-long range translational and orientational order. 

This work demonstrates the first appearance of an isotropic intermediate upon melting of a bosonic stripe phase and contributes to the growing literature on Bose liquid crystal analogues \cite{schmid_melting_2004, hickey_thermal_2014-1, bombin_berezinskii-kosterlitz-thouless_2019}. The data presented here suggest that the stripe-emulsion transition fits well into the broader context of a Kosterlitz-Thouless transition mediated by unbinding topological defects in 2D \cite{hammond_temperature_2005, toner_smectic_1981}; however, the finite-temperature transition is complicated by the presence of simultaneous dislocation defects in the stripe patterns, planar pseudospin vortices, and vortices in the superfluid phase. The precise interplay and role of each defect in the KT-like transition is a subject for future study.

This Letter provides insight into longstanding questions regarding the finite-temperature behavior of Rashba bosons and provides an explicit example of severely fragmented (quasi) condensation in a circular flat band, where the number of degenerate single-particle states is greater than the number of atoms in the system \cite{mueller_fragmentation_2006}. In this case, a singly condensed momentum state is absent and instead a spin-correlated, structured normal fluid phase emerges with a circular manifold of occupied momentum modes. This first computational prediction of an emulsion phase in cold atom systems serves to highlight a confluence among seemingly disparate fields -- soft matter physics, electronic condensed matter physics, and atomic physics. 

The authors thank Matthew Fisher and David Weld for helpful discussions. 
This work was enabled by field-theoretic simulation tools developed under support from the National Science Foundation (CMMT Program, DMR-2104255). Use was made of computational facilities purchased with funds from the NSF (CNS-1725797) and administered by the Center for Scientific Computing (CSC). This work made use of the BioPACIFIC Materials Innovation Platform computing resources of the National Science Foundation Award No. DMR-1933487. The CSC is supported by the California NanoSystems Institute and the Materials Research Science and Engineering Center (MRSEC; NSF DMR 1720256) at UC Santa Barbara. E.C.M acknowledges support from a Mitsubishi Chemical Fellowship, and L.B. acknowledges support from the DOE, Office of Science, Basic Energy Sciences under Award No. DE-FG02-08ER46524. 

\bibliography{Emulsion_paper_Bib}

\end{document}


\title{Supplementary Material: Emergence of a spin microemulsion in spin-orbit coupled Bose-Einstein condensates}
\author{Ethan C. McGarrigle} 
\email{emcgarrigle@ucsb.edu}
\affiliation{Department of Chemical Engineering, University of California, Santa Barbara, California 93106, USA}

\author{Kris T. Delaney} 
\affiliation{Materials Research Laboratory, University of California, Santa Barbara, California 93106, USA}
\author{Leon Balents}
 \affiliation{Kavli Institute for Theoretical Physics, University of California, Santa Barbara, California 93106, USA}
\affiliation{Canadian Institute for Advanced Research, Toronto, Ontario, Canada}
 
\author{Glenn H. Fredrickson}
\email{ghf@ucsb.edu}
\affiliation{Department of Chemical Engineering, University of California, Santa Barbara, California 93106, USA}
\affiliation{Materials Research Laboratory, University of California, Santa Barbara, California 93106, USA}
\affiliation{Department of Materials, University of California, Santa Barbara, California 93106, USA}


\maketitle

\section{Determining the Stripe to zFM Transition}  
The z-Ferromagnet (zFM) phase \cite{radic_stoner_2014} consists of two degenerate phases ($\ket{\uparrow}$ and $\ket{\downarrow}$ superfluids) that are macroscopically phase separated but in phase-coexistence equilibrium. In immiscible pseudospin mixtures where each particle number $N_{\alpha}$ is held fixed, there will be a sharp, dividing interface between the coexisting superfluids. Comparisons between a two-phase superfluid and a periodic stripe phase would be difficult in canonical ensemble simulations with particle number $N_{\alpha}$ fixed. Free energy comparisons would require large simulation cells to prove the two-phase superfluid lacks periodicity while showing diminishing free energy contributions from the interface. A more efficient method to probe the zFM phase is to use the grand canonical ensemble, where the chemical potential $\mu$ of each pseudospin species is equal and fixed. There, the system relaxes to a configuration that minimizes the natural thermodynamic potential — the grand potential $\Omega$ — by selecting either a pure $\ket{\uparrow}$ or $\ket{\downarrow}$ homogeneous superfluid state without any interface.

To study the stripe to zFM phase transition, we extend the methods of reference \cite{fredrickson_direct_2022} to calculate the grand potential for each phase in the metastable window where the zFM and stripe phases coexist. While thermodynamic potentials are cumbersome to access in particle simulations, field-theoretic representations provide direct access to operators that can be averaged in a field-theoretic simulation to yield free energy estimates \cite{fredrickson_direct_2022}. Here we detail how to estimate the grand potential in field theoretic simulations and provide a brief summary of our methods.

The grand potential is defined using standard thermodynamic relations \cite{fetter_quantum_2012} for a two-dimensional system with size (area) $A$: 
  \begin{equation}
    \begin{split}
        \Omega ( \mu, A, T) &= F - \mu N = E - TS - \mu \sum_{\alpha} N_{\alpha}, \\ 
         d\Omega &= -S dT - P dA - \sum_{\alpha} N_{\alpha} d\mu , \\ 
          P &= -\left (\frac{\partial \Omega}{\partial A} \right )_{\mu, T} ,
         \label{eq: Omega}
       \end{split}
   \end{equation}
  \noindent{where} $F$ is the Helmholtz free energy and $\mu$ is the chemical potential for both pseudospin boson species. For homogeneous systems, the grand potential is both extensive and a function of only one extensive quantity (system size), therefore $\Omega$ must be a homogeneous function of order 1 in the system size $A$ and can thus be expressed $\Omega (\mu, A, T) = A \frac{\partial \Omega}{\partial A}|_{T, \mu} $. From the differential form in equation\ (\ref{eq: Omega}), the area derivative is identified readily as the pressure, and the grand potential can be evaluated for a homogeneous system, in agreement with reference \cite{fetter_quantum_2012}: 
   \begin{equation} 
     \Omega = - PA.
      \label{eq: Omega_operator}
     \end{equation}   
Next, we extend the argument to the case of an ordered, \emph{periodic} mesophase, with direct application to the superfluid stripe phase in this work. For stripe systems that are scaled based on the unit cell that is commensurate with the stripe's periodicity, all intensive properties are invariant by definition. Therefore, the above arguments for a homogeneous system apply and $\Omega$ must be a homogeneous function of order 1 as long as the simulation cell size is commensurate with the stripe's periodicity $\frac{2 \pi n}{\tilde{\kappa}}$ for all positive integers $n$. As a result, we calculate $\Omega$ using equation\ (\ref{eq: Omega_operator}) for both a periodic stripe and a spatially homogeneous zFM phase. 

To link macroscopic thermodynamics with the microscopic details of the system, we utilize the definition of the grand potential from quantum statistical mechanics $\Omega (\mu, A, T) = - \frac{1}{\beta} \ln ( \mathcal{Z} )$ \noindent{where} $\mathcal{Z}$ denotes the grand canonical partition function and $\beta = 1 / k_{B} T$. Our simulations naturally represent conditions of fixed $\mu$ and $T$, so the pressure provides thermodynamic information related to the grand potential. With this link between the grand potential and the grand partition function, we determine a pressure operator functional in the grand canonical ensemble via the definition of pressure in equation\ (\ref{eq: Omega}): 
 \begin{equation}
      P  =  \frac{1}{\beta} \frac{\partial \ln (\mathcal{Z} )}{\partial A} = - \frac{1}{\beta} \left \langle \frac{\partial S[\phi, \phi^*]}{\partial A} \biggr |_{\mu , T} \right \rangle ,
    \label{eq: P}
  \end{equation} 
   \noindent{where} the area derivative is determined analytically using a procedure outlined in reference \cite{fredrickson_direct_2022}, yielding a pressure field operator functional:  
   \begin{widetext}
  \begin{align}
      \beta P[\bm{\phi}, \bm{\phi^*}] = \frac{-\tilde{\beta} }{N_{\tau} A} \sum_{j=0}^{N_{\tau}-1} \sum_{\alpha, \gamma} \int d^2 r \left [\phi^*_{\alpha, j} (\mathbf{r}) \frac{\partial \mathcal{\hat K}_{\alpha, \gamma}}{\partial A} \hspace{1.5px} \phi^{\vphantom{*}}_{\gamma, j-1} (\mathbf{r})  
         - \frac{\tilde{g} }{2} \phi^*_{\alpha, j} (\mathbf{r}) \phi^*_{\gamma, j} (\mathbf{r}) (\underline{\underline{I}} + \eta_{g} \underline{\underline{ \sigma}}^{x} )^{\vphantom{*}}_{\alpha \gamma} \phi^{\vphantom{*}}_{\gamma, j-1} (\mathbf{r}) \phi^{\vphantom{*}}_{\alpha, j-1} (\mathbf{r}) \right ] , 
    \label{eq: P_op}
    \end{align} 
    \end{widetext}  
  \begin{equation}
   \frac{\partial \underline{\underline{ \mathcal{\hat K} }}}{\partial A} = 
   \begin{bmatrix}
       \tilde{\nabla}^2  & - \tilde{\kappa} [i \partial_{\tilde{x}} + \partial_{\tilde{y}}] \\ 
       - \tilde{\kappa} [i \partial_{\tilde{x}} - \partial_{\tilde{y}}]   &  \tilde{\nabla}^2  
       \end{bmatrix} .
  \end{equation} 
From a thermal-averaged pressure, we determine a dimensionless grand potential using $\beta \Omega = - \langle \beta P \rangle A$, where the thermal average is evaluated using a sample average over Langevin time as described in the \textit{Methods} of the main text. 

\begin{figure}[ht] 
 \includegraphics[scale = 0.32]{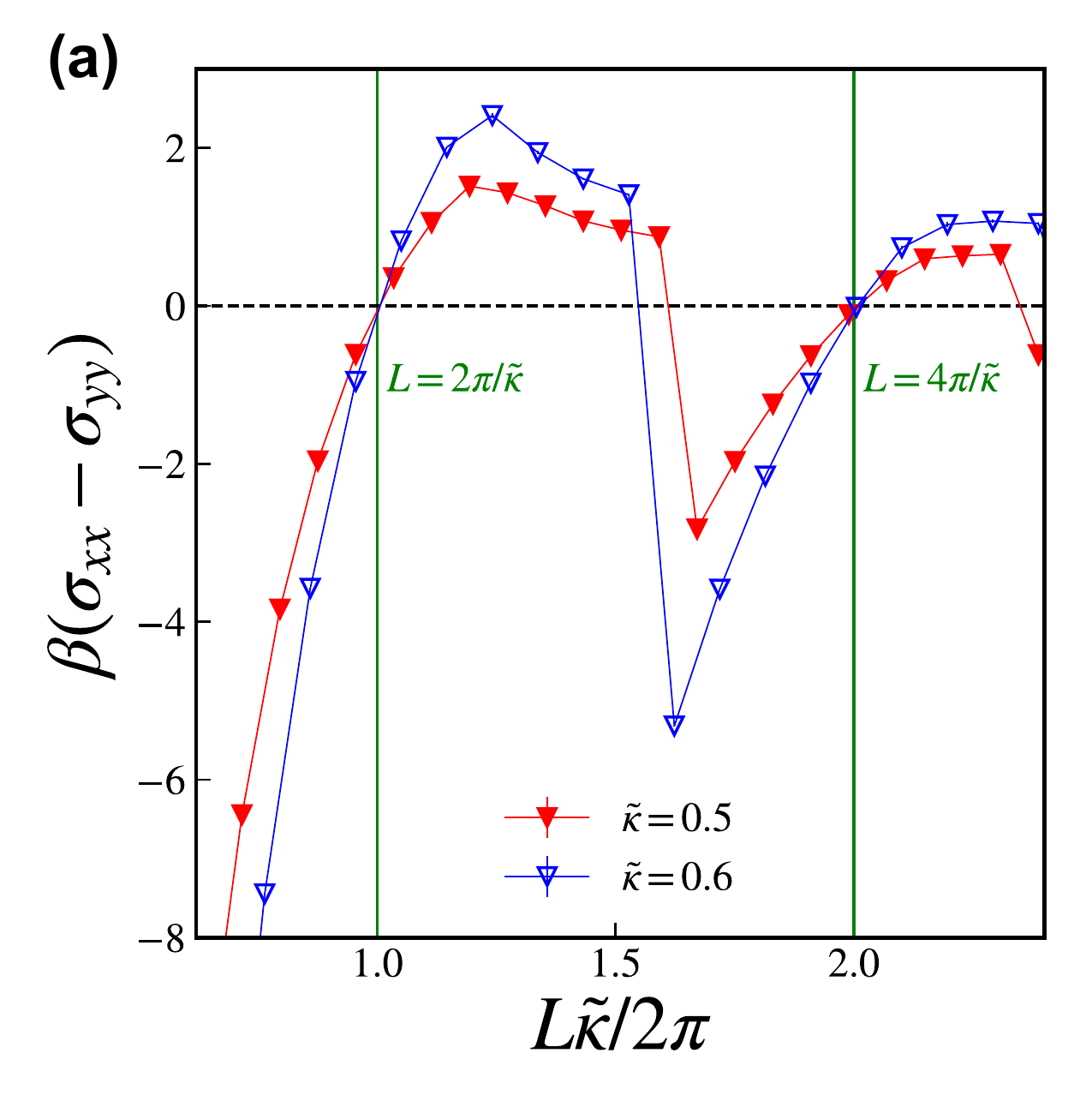} \label{fig: SI1a} 
 \includegraphics[scale = 0.32]{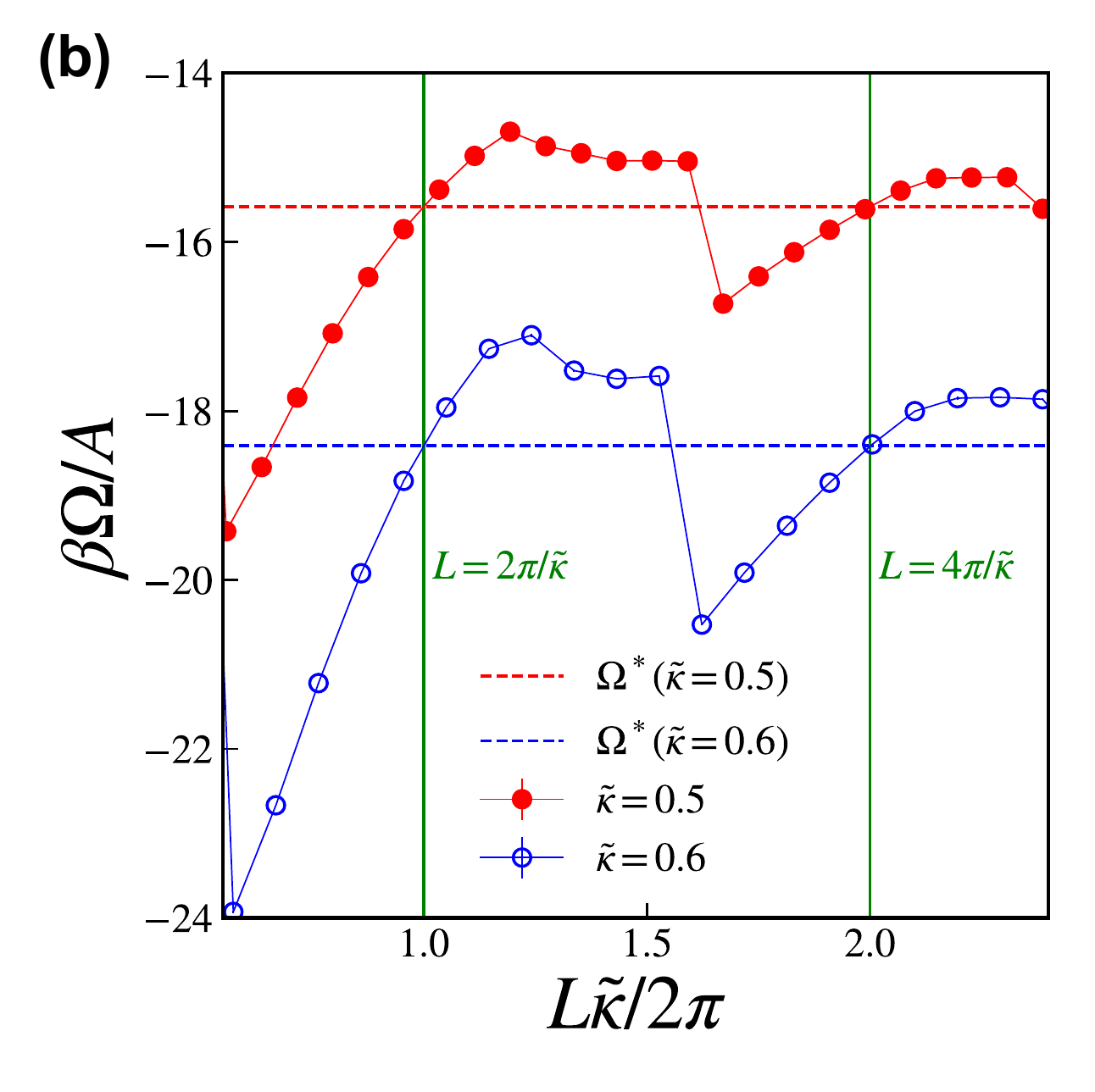} \label{fig: SI1b}
 \caption{Stripe phase commensurability analysis for $\tilde{T} = 1.0$, $\eta_{g} = 1.1$, $\tilde{g} = 0.05$, and two SOC coupling strengths $\tilde{\kappa} = 0.5$ and $\tilde{\kappa} = 0.6$ . (a) Difference in cell stress tensor diagonal components for different simulation cell sizes $L$. Isotropic stress conditions occur when the diagonal stress difference intersects the dashed black line, which corresponds precisely to $L^* = \frac{2 \pi n}{\tilde{\kappa}}$.  (b) Intensive grand potential as a function of system size. The intensive grand potential $ \beta \Omega^* (\tilde{\kappa}) / A $ values for $\tilde{\kappa} = 0.5$ and $\tilde{\kappa} = 0.6$ at commensurate conditions are shown with dashed horizontal lines. Sharp jumps occur near $L \tilde{\kappa} / 2 \pi = 1.5$ and correspond to the system adding another set of stripes to accommodate the larger cell size. Error bars are standard errors of the mean determined in the Langevin sampling averaging process and are smaller than the symbol size.}
\label{fig: SI1}
\end{figure} 

Before comparing the grand potentials of each phase, we performed a commensurability analysis to ensure the stripe structures were commensurate with the simulation cell. Commensurability effects are significant for the stripe phase, which has liquid crystalline-like order and shows significant free energy deviations away from commensurate cell conditions (see Figure\ (\ref{fig: SI1}b)). The microemulsion phase is much softer and does not show significant free energy sensitivity to cell size. 

\begin{figure}[ht] 
	\includegraphics[scale = 0.30]{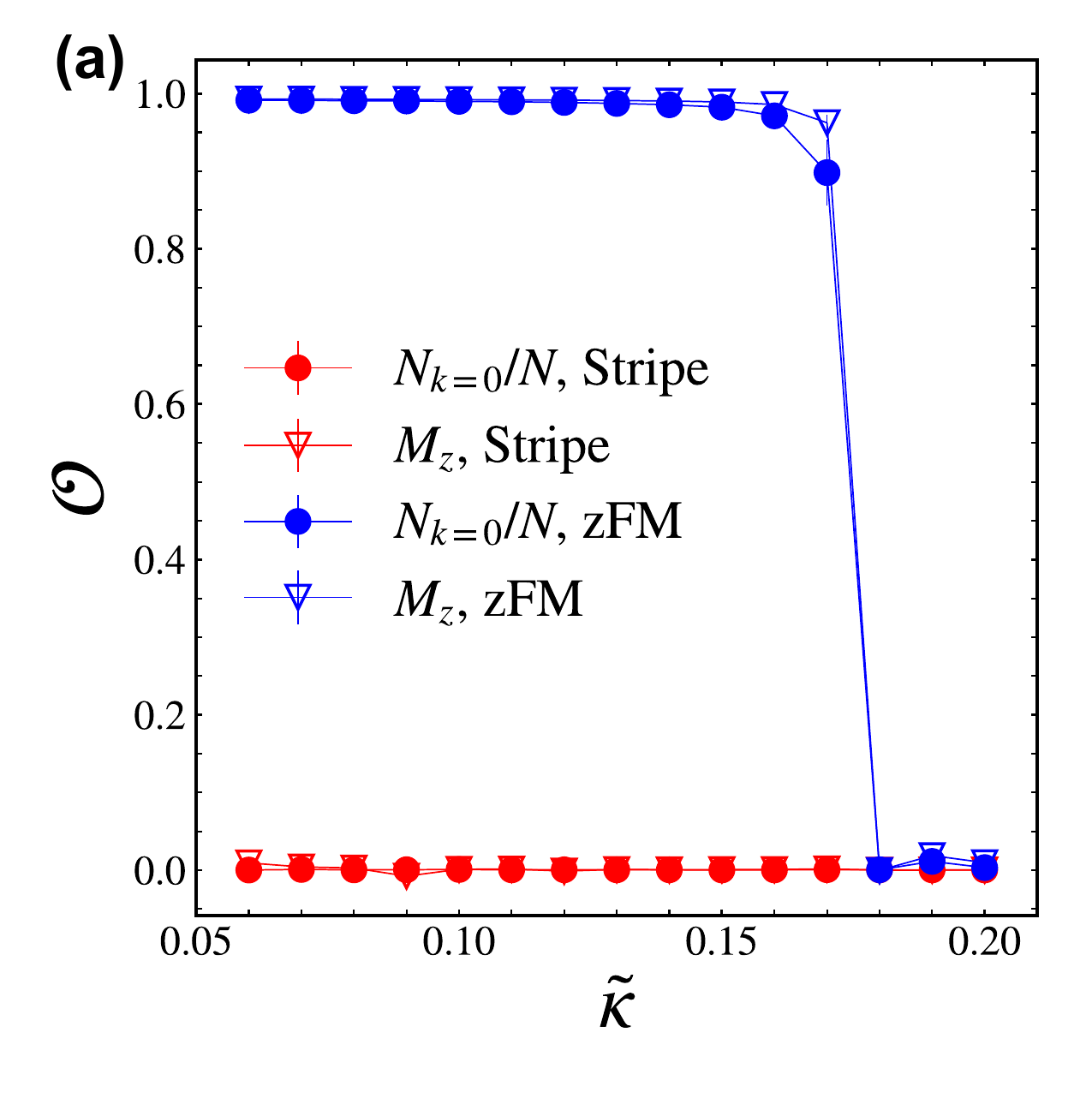} 
	\includegraphics[scale = 0.30]{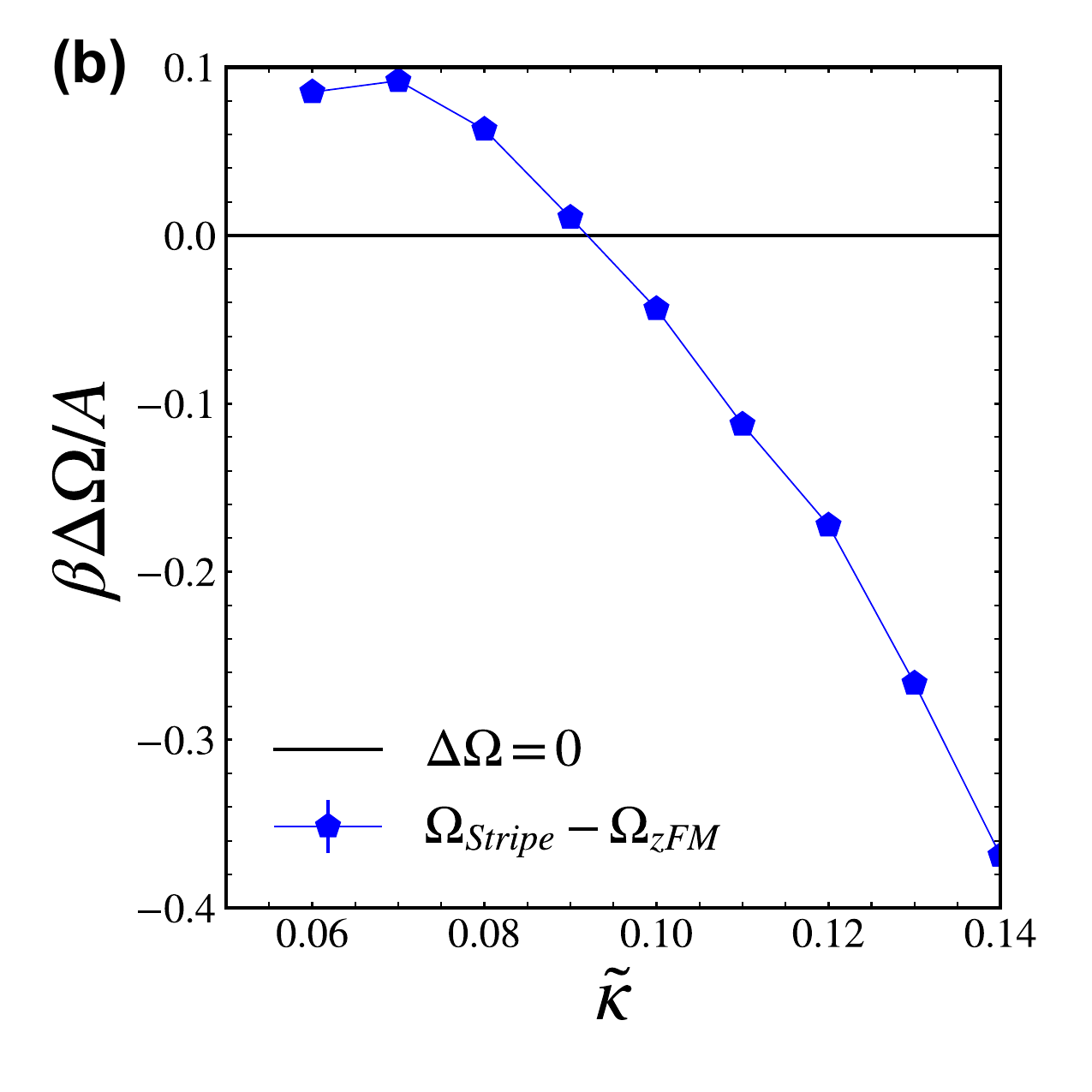}  
           \includegraphics[scale = 0.36]{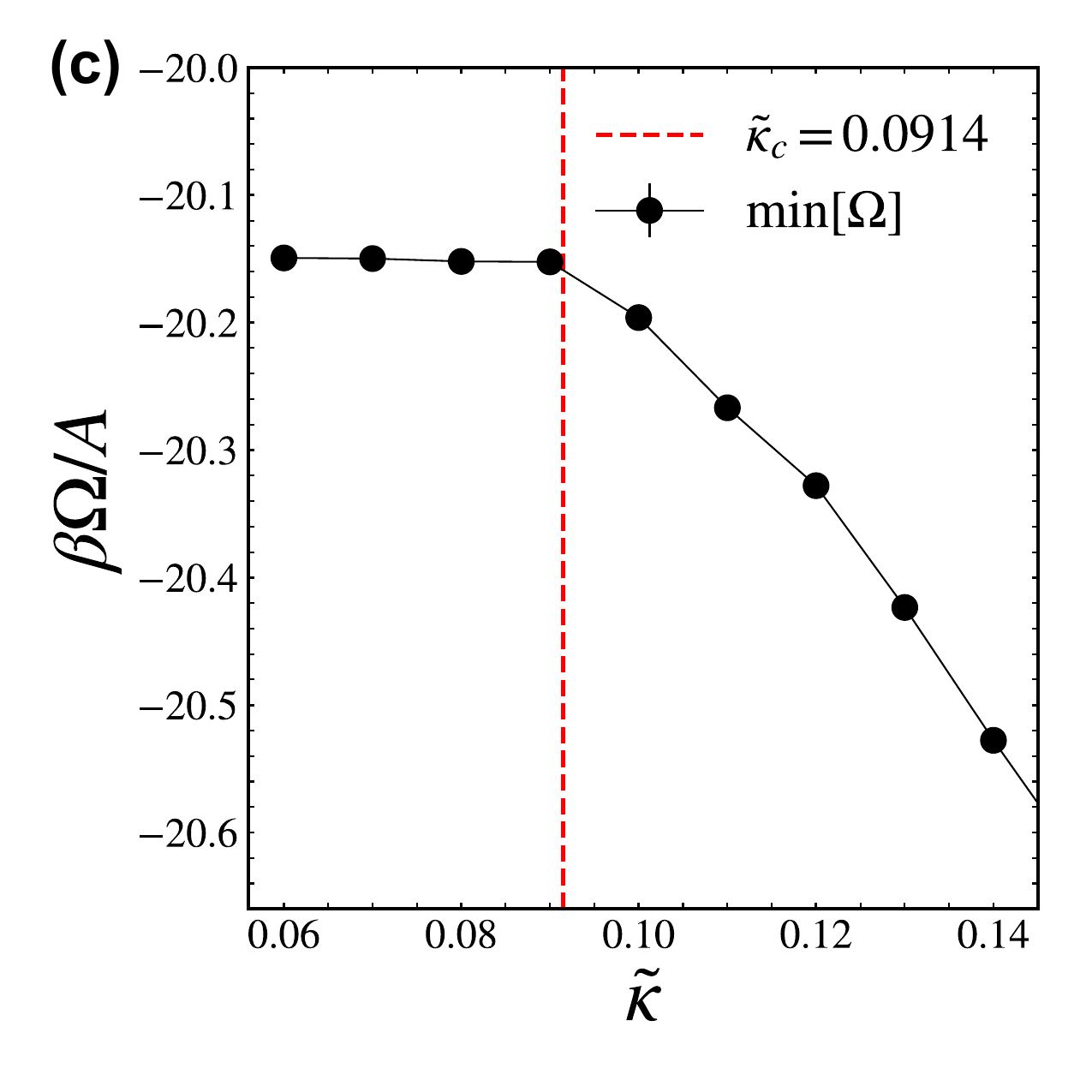}   
   \caption{ First order stripe to zFM transition at $\tilde{T} = 0.5$, $\eta_{g} = 1.1$, $\tilde{g} = 0.05$.  (a) The $\hat{z}$ magnetization and $\mathbf{k}=\mathbf{0}$ occupation fraction plotted at different $\tilde{\kappa}$ for the stripe and zFM phases. The zFM phase becomes unstable near $\tilde{\kappa} > 0.17$, where the order parameters jump from 1 to 0, characteristic of a first order transition. (b) Intensive grand potential difference between the zFM and stripe phases. The crossing point with the solid black line determines the SOC strength $\tilde{\kappa}_{c}$ at the phase boundary. (c) Grand potential $\Omega (\tilde{\kappa})$ plotted using values from the phase with the lower grand potential. The kink is evident near $\tilde{\kappa}_{c}$, shown with the red dashed line. For all plots, error bars are standard errors of the mean determined from the complex Langevin data.}  
  \label{fig: SI2}
\end{figure}

The commensurate cell size was determined using a cell stress analysis following the method outlined in reference \cite{fredrickson_direct_2022} for structured fluids, and an example is shown in Figure\ (\ref{fig: SI1}). An excess cell stress operator is derived by quantifying changes in the intensive grand potential with respect to the cell dimensions at constant chemical potential and temperature: 
$$ \beta \underline{\underline{\sigma}} =  \frac{1}{A} \underline{\underline{h}} \frac{\partial ( \beta \Omega_{\text{ex}}) }{\partial \underline{\underline{h}}}\biggr |_{\mu, T}  = \frac{ 1 }{A} \underline{\underline{h}} \left \langle \frac{\partial S}{\partial \underline{\underline{h}}}\biggr |_{\mu, T} \right \rangle  ,  $$ 
where $\Omega_{\text{ex}}$ is the excess grand potential and $\underline{\underline{h}}$ is a symmetric and diagonal cell tensor with $h_{xx} = L_{x}$, $h_{yy} = L_{y}$ and $h_{xy} = h_{yx} = 0$. The diagonal components of the dimensionless excess stress in the $\nu$ direction are  
   \begin{widetext}
    \begin{align}
      \beta \sigma_{\nu \nu} [\bm{\phi}, \bm{\phi^*}]  = \frac{2 \tilde{\beta} }{N_{\tau} A} \sum_{j=0}^{N_{\tau}-1} \sum_{\alpha, \gamma} \int d^2 r \left [ \phi^*_{\alpha, j} (\mathbf{r}) \frac{\partial \mathcal{\hat K}_{\alpha, \gamma}}{\partial L_{\nu} } \hspace{0.5px} \phi^{\vphantom{*}}_{\gamma, j-1} (\mathbf{r}) 
         - \frac{\tilde{g} }{4} \phi^*_{\alpha, j} (\mathbf{r}) \phi^*_{\gamma, j} (\mathbf{r}) (\underline{\underline{I}} + \eta_{g} \underline{\underline{ \sigma}}^{x} )^{\vphantom{*}}_{\alpha \gamma} \phi^{\vphantom{*}}_{\gamma, j-1} (\mathbf{r}) \phi^{\vphantom{*}}_{\alpha, j-1} (\mathbf{r}) \right ],  
    \label{eq: Stress_op}
      \end{align}
    \end{widetext}
    \vskip -2ex
   \begin{equation}
   \frac{\partial \underline{\underline{ \mathcal{\hat K} }}}{\partial L_{x}} = 
   \begin{bmatrix}
       \partial^2_{\tilde x}  & - i \tilde{\kappa} \partial_{\tilde{x}}  \\
       -i \tilde{\kappa} \partial_{\tilde{x}}   &  \partial^2_{\tilde x}   
       \end{bmatrix} \nonumber
  \hskip 10ex
   \frac{\partial \underline{\underline{ \mathcal{\hat K} }}}{\partial L_{y}} = 
   \begin{bmatrix}
       \partial^2_{\tilde y}  & - \tilde{\kappa} \partial_{\tilde{y}}  \\
       \tilde{\kappa} \partial_{\tilde{y}}   &  \partial^2_{\tilde y}   
       \end{bmatrix} .
\end{equation}

The cell stress tensor $\underline{\underline{\sigma}}$ becomes isotropic at the commensurate cell size for periodic stripe phases, so we vary the cell length $L$ for vertically oriented stripe structures and monitor the cell stress tensor's diagonal components $\sigma_{x x}$ and $\sigma_{y y}$. We find in Figure\ (\ref{fig: SI1}a) that the commensurate cell length is $L^{*} = 2 \pi n / \tilde{\kappa} $ for a choice of $\tilde{\kappa}$. Figure\ (\ref{fig: SI1}b) shows that the intensive grand potential is invariant to the number of stripes $n$ at the commensurate cell conditions $L = L^*$. For all simulations of the stripe superfluid, we use a square cell $(L_{x} , L_{y} ) = ( L^* , L^* )$ that is commensurate with a vertical stripe structure determined by this analysis. 
 
The phase transition between a finite-temperature stripe and zFM phase is classified as first order and shown in Figure\ (\ref{fig: SI2}). The zFM is a spatially homogeneous, superfluid phase with ferromagnetic $\hat{z}$ pseudospin order, so the $\mathbf{k} = \mathbf{0}$ momentum state occupation fraction and area-averaged $\hat{z}$ magnetization $M_{z}$ serve as good order parameters to show the presence of homogeneous superfluidity and $\hat{z}$ magnetic order, respectively. In contrast, the vertically-oriented stripe superfluid has two macroscopically occupied modes at non-zero $\mathbf{k} = (\pm \tilde{\kappa}, 0)$ and no net $\hat{z}$ ferromagnetism. Figure\ (\ref{fig: SI2}a) shows discontinuities in each order parameter near $\tilde{\kappa} = 0.17$ at $\tilde{T} = 0.5$ for the zFM phase and signals the closure of the zFM-stripe metastability window. The stripe structures remain stable down to low spin-orbit coupling (SOC) strengths and may require prohibitively large simulation cells to observe stripe structure instabilities as $\tilde{\kappa} \to 0$. 

 For this first-order transition, we utilize the grand potential $\Omega$ to locate the precise boundary between the stripe and zFM phase at the prescribed temperature $\tilde{T}$ and chemical potential $\mu$.  The phase with the lower grand potential is the stable equilibrium phase, so the SOC strength at the phase boundary $\tilde{\kappa}_{c}$ is determined by finding where the grand potential difference between the competing phases crosses zero. A representative phase boundary calculation is shown in Figure\ (\ref{fig: SI2}b) for $\tilde{T} = 0.5$ as an example. The crossing point was determined by linear regression, where the propagated error and covariance of the fitted slope and intercept provided an estimation of error for $\tilde{\kappa}_{c}$. For the example shown in Figure\ (\ref{fig: SI2}) at $\tilde{T} = 0.5$, the phase boundary point was determined to be $\tilde{\kappa}_{c} = 0.0914 \pm 0.006 $, where a kink in $\Omega$ is observed (Figure (\ref{fig: SI2}c)).   

\section{Stripe to Emulsion Transition: Kosterlitz--Thouless Finite-Size Analysis} 

Near the critical temperature of the stripe to microemulsion transition, significant finite-size effects occur in estimates of the superfluid density $\rho_{\text{SF}}$, which obscure the value of the critical temperature.  In 2D superfluid systems, a Kosterlitz--Thouless (KT) transition is expected and is characterized by a universal jump in the superfluid density (stiffness) \cite{nelson_universal_1977} from a value of $\rho_{\text{SF}} (\tilde{T}_{\text{KT}}) = \frac{1}{\pi} \tilde{T}_{\text{KT}}$ to zero as the temperature is increased across $\tilde{T}_{\text{KT}}$, expressed using the dimensionless quantities employed in this study. However, the stripe to emulsion transition is broadened by the finite size of the simulation cell, which cannot capture the diverging correlation length and show a true jump in $\rho_{\text{SF}}$. As the system size increases, the decline of the superfluid stiffness sharpens in character to align with a universal jump. 

To calculate the superfluid density tensor, we use the phase twist method \cite{subasi_quantum-geometric_2022} described in the main text. From the fundamental relation $\Omega = -k_{B} T \ln (\mathcal{Z})$, the superfluid density tensor becomes a functional of coherent state field configurations that can be averaged across a Langevin simulation:
\begin{equation}
   \rho^{\mu \nu}_{\text{SF}}  = \langle \rho \rangle \delta_{\mu \nu} - \frac{2 \tilde{\beta}}{A} \left [ \langle \mathcal{P}_{\mu}  \mathcal{P}_{\nu} \rangle  - \langle \mathcal{P}_{\mu} \rangle \langle \mathcal{P}_{\nu} \rangle \right ] ,
 \end{equation} 
 \noindent{where} $\delta_{\mu \nu}$ is the Kronecker delta, $\rho = (N_{\uparrow} + N_{\downarrow}) / A $ is the total particle number density, $A$ is the system size, $\tilde{\beta}$ is the dimensionless inverse temperature, and $\mathcal{P}_{\nu}$ denotes the dimensionless physical momentum field operator in the $\nu$ direction, which is a functional of the coherent state fields: 
  \begin{equation}
    \begin{split}
     \mathcal{P}_{\nu} [\bm{\phi}, \bm{\phi}^*] &= \frac{1}{N_{\tau}} \sum_{j=0}^{N_{\tau} - 1} \sum_{\alpha} \int d^2 r \hspace{2px} \phi^*_{\alpha, j} (\mathbf{r}) [ -i \partial_{\tilde{\nu}} ] \phi^{\vphantom{*}}_{\alpha, j-1} (\mathbf{r}) \\
     &- \tilde{\kappa} \int d^2 r \hspace{2px} M_{\nu} [\bm{\phi}^{\vphantom{*}}, \bm{\phi}^* ; \mathbf{r}],
    \end{split}
  \end{equation} 
 \noindent{where} we have utilized the pseudo-magnetization field operator defined in the main text. The superfluid density is a rank-2 tensor in $d=2$ and obeys mass conservation $\rho = \rho_{\text{NF}} + \rho_{\text{SF}}$, where $\rho_{\text{SF}} = \frac{1}{2} \text{Tr}[\underline{\underline{\rho}}_{\text{SF} }]$, where $\rho_{\text{NF}}$ and $\rho_{\text{SF}}$ are the normal and superfluid number densities, respectively. 

\begin{figure}[ht] 
      \includegraphics[scale = 0.28]{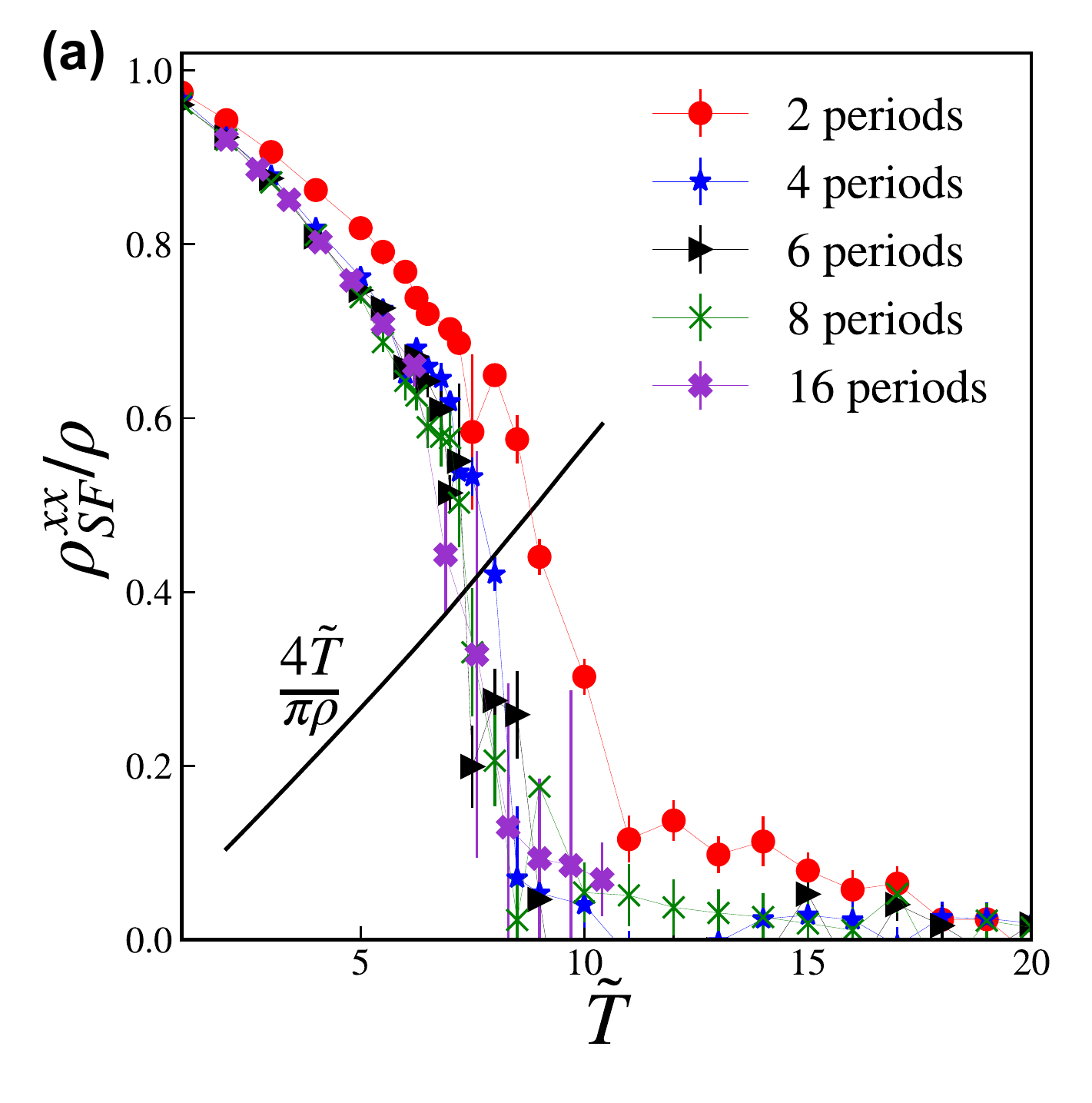} 
   \includegraphics[scale = 0.32]{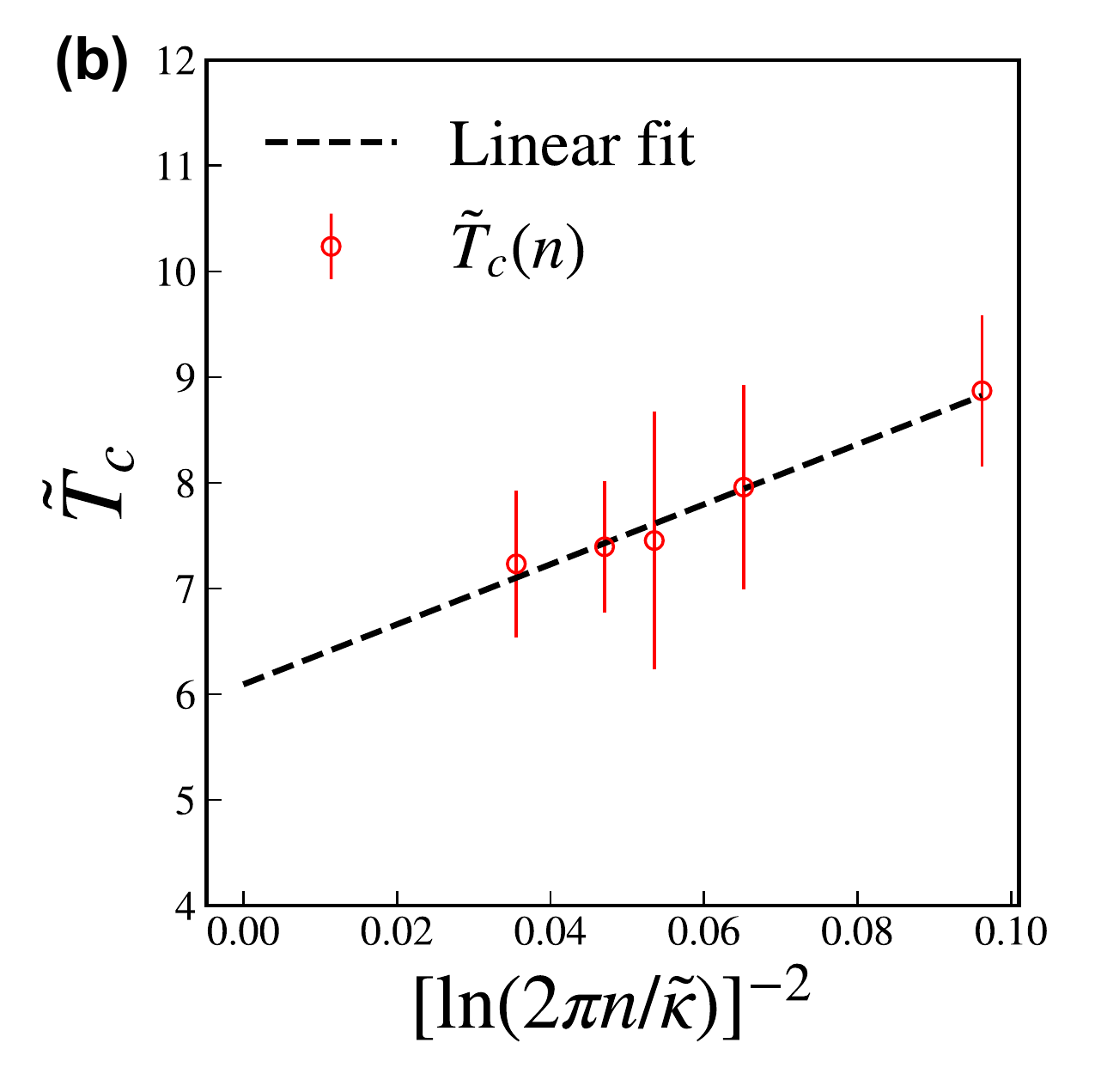} \label{fig: SI3b}
  \caption{Kosterlitz--Thouless-like transition at $\tilde{\kappa} = 0.5$, $\eta_{g} = 1.1$, and $\tilde{g} = 0.05$. (a) Superfluid fraction normal component $ \rho^{xx}_{\text{SF}} / \rho$ plotted for several system sizes, denoted as the number of stripe periods $n$ in the legend. Error bars are standard errors of the mean, determined from the Langevin-time averaging process. The nematic superfluid universal jump result is shown in the solid black curve. (b) $\tilde{T}_{c}$ scaling with system size using the scaling form in Equation\ (\ref{eq: T_KT_scaling}), with a linear fit shown in the black dashed line. Error bars for each $\tilde{T}_{c} (n)$ were estimated from the slope and intercept variance of a crossing between the $\rho^{xx}_{\text{SF}} (n) / \rho $ curve and the modified universal jump via linear regression. }
    \label{fig: SI3}
    \end{figure}

 In the stripe to microemulsion transition, the superfluid stiffness component in the stripe layer normal direction $\rho^{xx}_{\text{SF}}$ was used as the representative order parameter to perform finite-size scaling analysis. Although the superfluid stiffness parallel to the stripes $\rho^{yy}_{\text{SF}}$ shows a similar decline, the normal component $\rho^{x x}_{\text{SF}}$ experiences a full variation between 0 and 1 and enables for a more accurate finite-size analysis. Data are plotted for $\tilde{\kappa} = 0.5$ in Figure\ (\ref{fig: SI3}a), where the decline in the superfluid density is steeper as the number of stripe periods increases. All stripe simulations used initial configurations with stripes oriented with normals along the $\hat{x}$ direction to ensure consistency in this anisotropy, and the time scale for rotations in stripe orientation exceed the simulation time. 
 
\begin{figure}[ht] 
    \includegraphics[scale = 0.29]{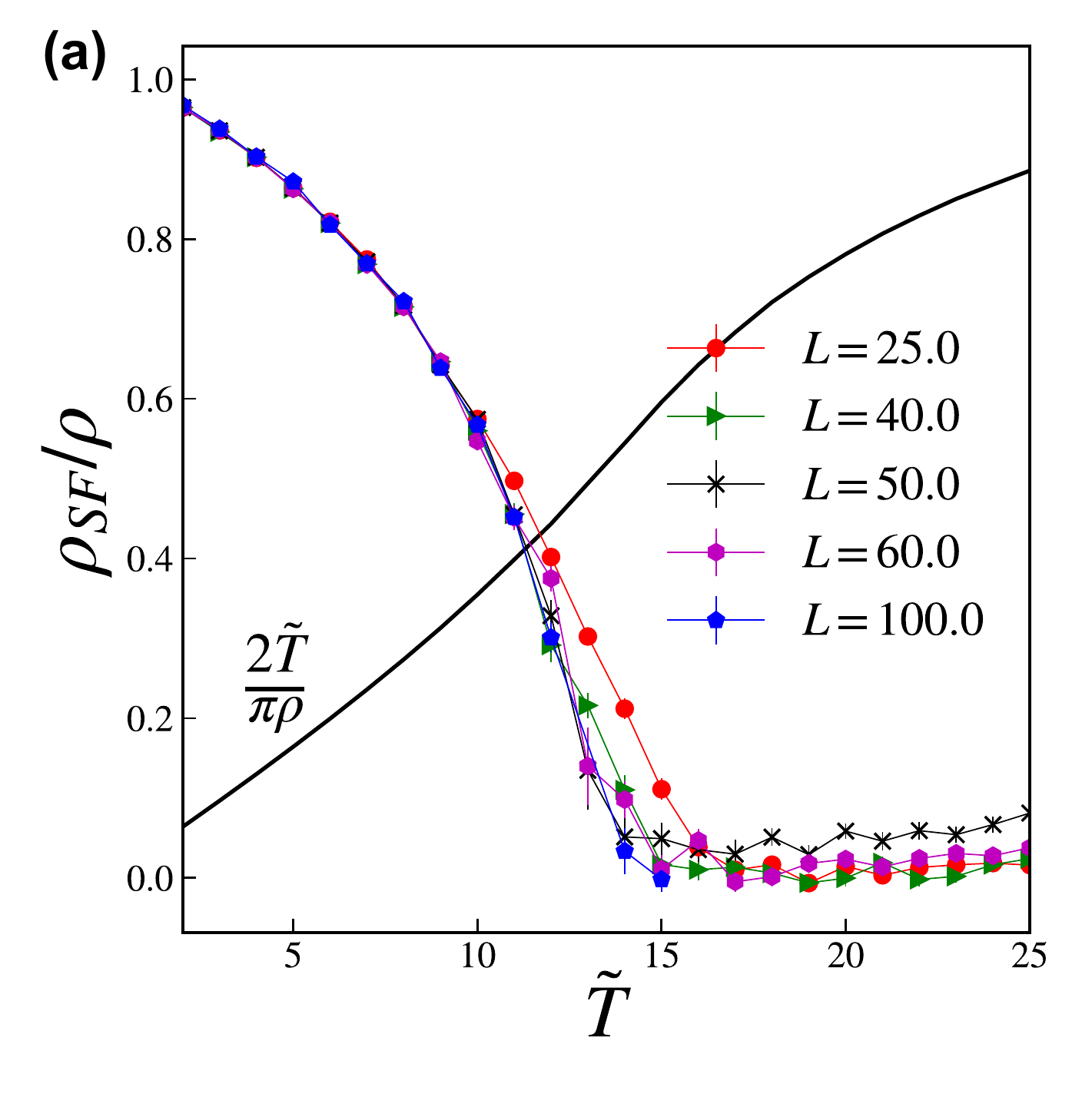} 
   \includegraphics[scale = 0.31]{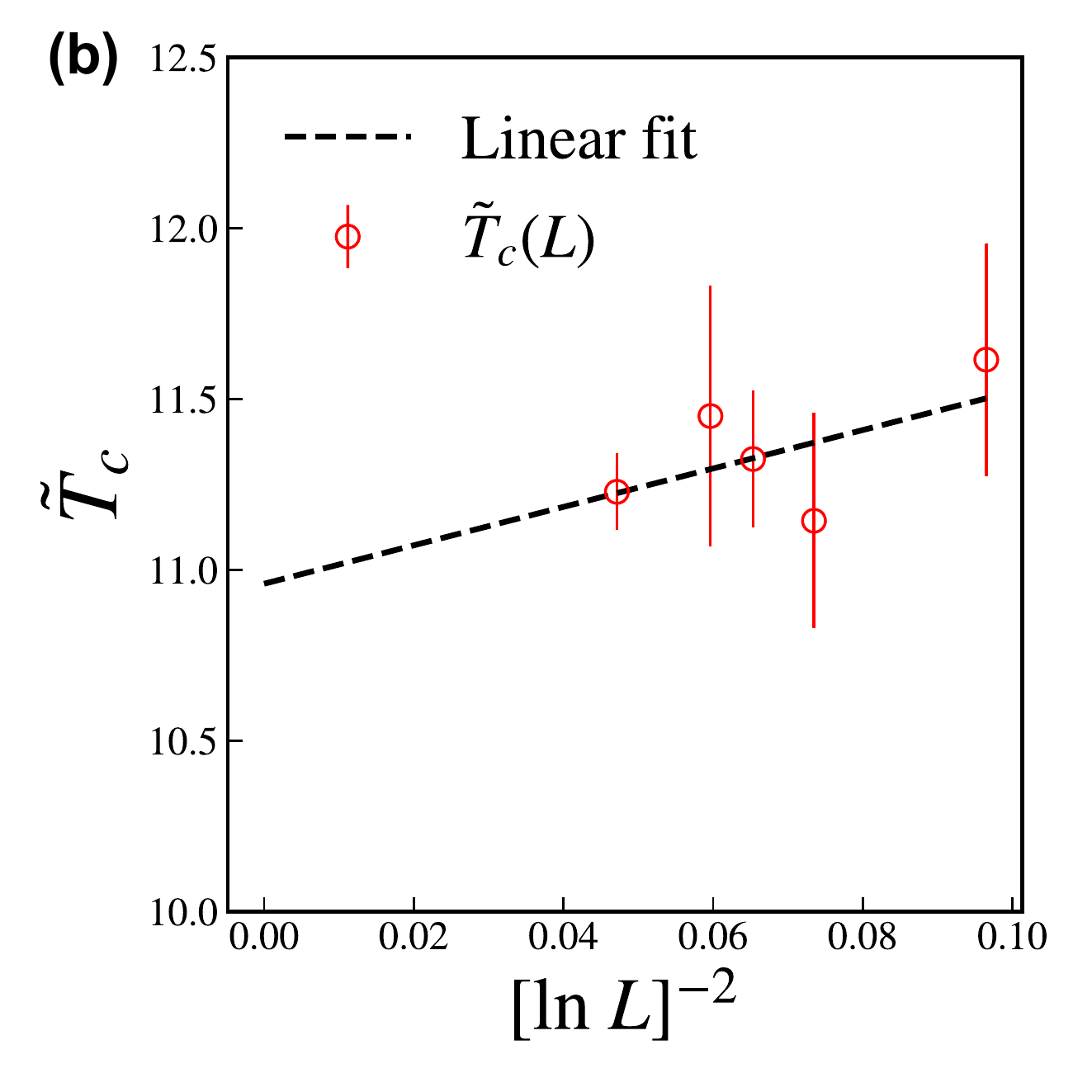} 
 \caption{ Kosterlitz--Thouless-like transition at $\tilde{\kappa} = 0.0$, $\eta_{g} = 1.1$, and $\tilde{g} = 0.05$. (a) Superfluid fraction $ \rho_{\text{SF}} / \rho $ of the zFM phase plotted for several system sizes with length $L$. Error bars are standard errors of the mean, determined from the Langevin-time averaging process. The modified universal result for pseudospin-1/2 Bose gases is shown in the solid black curve. (b) Scaling of the critical temperature with system size. Open red circles are determined from the crossing of the solid black curve $\rho_{\text{SF}} (\tilde{T}_{\text{KT}}) = \frac{2}{\pi} \tilde{T}_{\text{KT}}$ with superfluid density data at each system size. Points are plotted using the scaling form in Equation\ (\ref{eq: T_KT_scaling}), with a linear fit shown in the black dashed line. Error bars for each $\tilde{T}_{c} (L)$ estimate were estimated from the slope and intercept variance of the crossing between the $\rho_{\text{SF}} (L) / \rho $ curve and the modified universal jump via linear regression. } 
 \label{fig: SI4}
\end{figure}
 
For multi-species interacting Bose-Einstein condensates, the jump in the superfluid density is reported to deviate from the universal value \cite{kobayashi_berezinskii-kosterlitz-thouless_2019, hickey_thermal_2014-1, mukerjee_topological_2006}, so careful consideration must be made in our finite-size analysis. For example, reference \cite{kobayashi_berezinskii-kosterlitz-thouless_2019} showed that pseudospin-1/2 bosons without spin-orbit coupling ($\tilde{\kappa} = 0$) are expected to experience a jump at twice the universal result: $\rho_{\text{SF}} (\tilde{T}_{\text{KT}}) = \frac{2}{\pi} \tilde{T}_{\text{KT}}$. We follow reference\ \cite{jian_paired_2011}, which estimated the modified universal jump as $\rho_{\text{SF}} (\tilde{T}_{\text{KT}}) = \frac{4}{\pi} \tilde{T}_{\text{KT}}$ for the stripe phase with isotropic SOC, which is a nematic superfluid in the thermodynamic limit. Furthermore, reference \cite{mukerjee_topological_2006} showed that a nematic superfluid phase in $S=1$ spinor condensates are best described by a KT transition with a similarly modified universal jump $\rho_{\text{SF}} (\tilde{T}_{\text{KT}}) = \frac{4}{\pi} \tilde{T}_{\text{KT}}$. The precise nature of the bound topological defects dictates the adjustment to the precise temperature at which the universal jump occurs. As such, a detailed analysis of the interplay of superfluid phase, XY spin, and $\hat{z}$ spin defects is left to a future study. 

Following procedures outlined in references \cite{kobayashi_berezinskii-kosterlitz-thouless_2019, hickey_thermal_2014-1}, we estimate a corrected Kosterlitz--Thouless critical temperature $T_{\text{KT}}$ for systems with SOC ($\tilde{\kappa} > 0$) using the modified universal result. We extrapolate from the approximate critical temperatures at different system sizes, obtained from the crossing between the modified universal form $\rho_{\text{SF}} (\tilde{T}_{\text{KT}}) = \frac{4}{\pi} \tilde{T}_{\text{KT}}$ and the $\hat{x}$ superfluid stiffness curves shown in Figure\ (\ref{fig: SI3}a). The crossing temperature at each system size was determined using linear interpolation.  For systems without SOC ($\tilde{\kappa} = 0$), the other modified universal result $\rho_{\text{SF}} (\tilde{T}_{\text{KT}}) = \frac{2}{\pi} \tilde{T}_{\text{KT}}$ for pseudospin-1/2 Bose gases was employed to produce the results shown in Figure\ (\ref{fig: SI4}), corresponding to the unbinding of half-vortex defects.   

In a KT transition, finite-size corrections to the precise critical temperature are assumed to scale logarithmically in the following form: 
   \begin{equation}
    \tilde{T}_{c} (L) = \tilde{T}_{\text{KT}} + \frac{b}{ [\ln (L) ]^2 } 
     \label{eq: T_KT_scaling}
   \end{equation} 
\noindent{where} $L$ is the linear system size (simulation cell box length) and $b$ is an arbitrary, non-universal parameter. The system size replaces the growing bulk correlation length near the critical transition. We assume such a logarithmic form for our finite-size analysis. Figures\ (\ref{fig: SI3}b) and (\ref{fig: SI4}b) shows a least squares linear fit of the $\tilde{T}_{c} (L)$ data as a function of inverse system size as described in equation\ (\ref{eq: T_KT_scaling}). The intercept of the linear fit provides an estimate of the true Kosterlitz-Thouless transition temperature that corrects for finite-size errors. In the examples shown above, we estimate $\tilde{T}_{\text{KT}} = 6.13 \pm 0.96$ and $\tilde{T}_{\text{KT}} = 10.96 \pm 0.36$ for $\tilde{\kappa} = 0.5$ and $\tilde{\kappa} = 0$, respectively. $\tilde{T}_{\text{KT}}$ error estimates are determined via the intercept's variance from the linear fit to equation\ (\ref{eq: T_KT_scaling}).
   
\section{Microemulsion to Homogeneous Fluid Continuous Crossover} 
The transition between the structured normal fluid with microemulsion character and homogeneous normal fluid is best characterized as a continuous crossover. To quantify the crossover temperature, we monitored two features of the data: 1) The 2D isothermal compressibility $\kappa_{T} = -\frac{1}{A} \frac{dA}{dP}|_{T}$ , and 2) a homogeneous order parameter, calculated as the fraction of $\mathbf{k}=\mathbf{0}$ mode occupation relative to the maximally occupied momentum state $\frac{N(\mathbf{k} = \mathbf{0})}{\max_{\mathbf{k}} N(\mathbf{k})}$ in the distribution. This order parameter is unity in the spatially homogeneous superfluid and normal fluid and trends to zero deep in the microemulsion where the $\mathbf{k} \neq \mathbf{0}$ ring modes are predominantly populated. 

In the grand canonical ensemble, the isothermal compressibility can be readily accessed via the population variance of the total particle number \cite{fredrickson_field-theoretic_2023}: 
   \begin{equation}
   \kappa_{T}  =  \beta A \left [ \frac{ \langle N^2 \rangle }{\langle N \rangle^2} - 1 \right ] 
 \end{equation} 
 \noindent{where} the total particle number $N$ can be accessed via spatial integration of the density field operator defined in the main text:  $ N[\bm{\phi}, \bm{\phi}^*] = \sum_{\alpha} \int d^2 r \hspace{2px} \rho_{\alpha} [\bm{\phi}, \bm{\phi}^* ; \mathbf{r} ] $. 
\begin{figure}[htp] 
    \includegraphics[scale = 0.375]{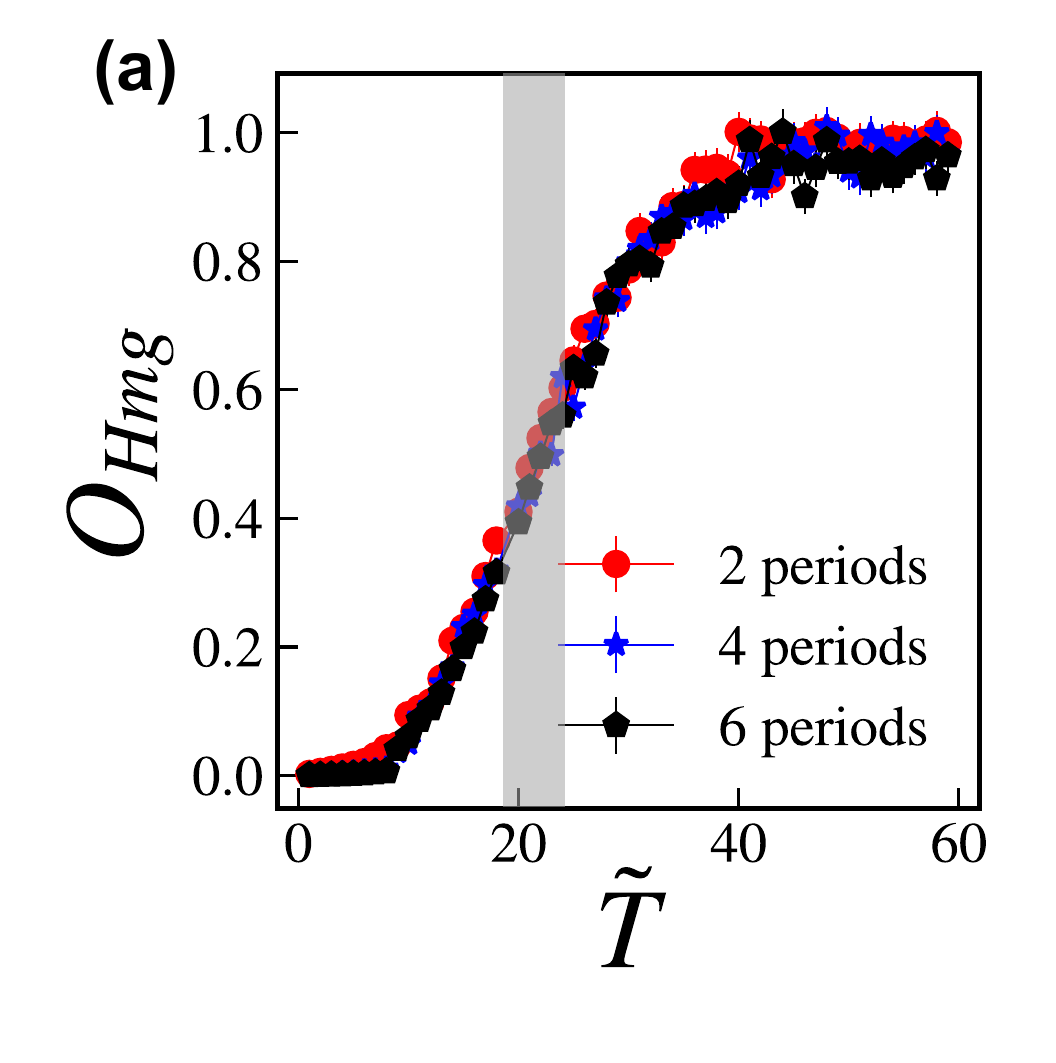} 
   \includegraphics[scale = 0.37]{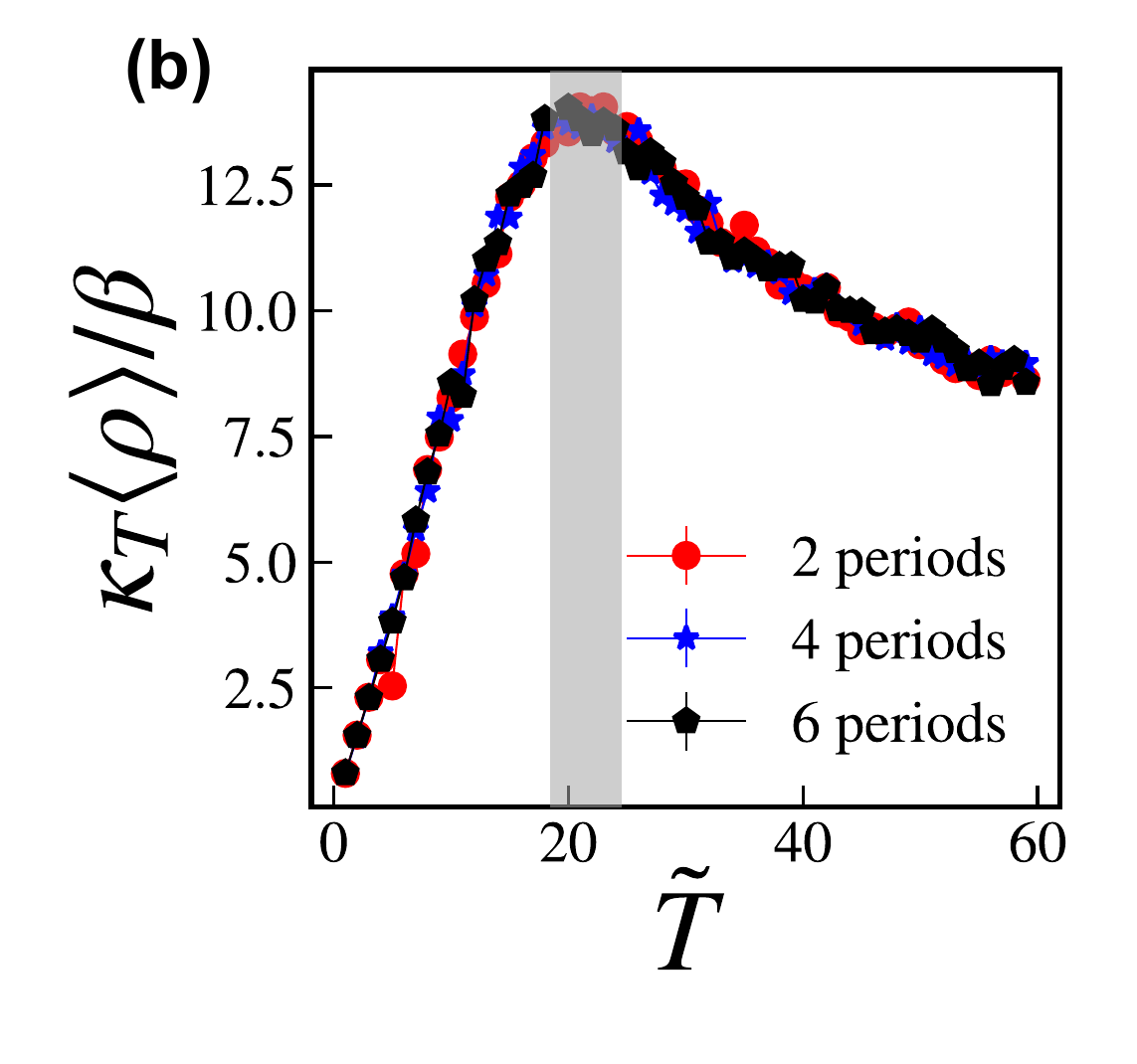}
   \caption{ Continuous crossover at $\tilde{\kappa} = 0.6$, $\eta_{g} = 1.1$, and $\tilde{g} = 0.05$. (a) Homogeneous order parameter throughout the melting process. Error bars are standard errors of the mean. (b) Isothermal compressibility, rescaled for better curve collapse.  In both plots, error bars are standard errors of the mean.} 
    \label{fig: SI5}
\end{figure}

 Figure\ (\ref{fig: SI5}) presents data for $\tilde{\kappa} = 0.6$ as an example, showing the continuous crossover from the microemulsion to homogeneous fluid. As temperature increases after the stripe phase melts, the homogeneous order parameter increases monotonically (Figure\ (\ref{fig: SI5}a)) and saturates once the fluid is completely homogeneous. We fit the homogeneous order parameter data to a sigmoid and extract the crossover temperature $\tilde{T}_{\text{crossover}}$ as the approximate centroid of the temperature data with an estimated error derived from the covariance of the sigmoid's centroid. For this example, we determine the crossover temperature to be $\tilde{T}_{\text{crossover}} = 21.1 \pm 0.1$. This $\tilde{T}_{\text{crossover}}$ value showed excellent agreement with the approximate position of the peak in the compressibility at all system sizes, shown in Figure\ (\ref{fig: SI5}b). In both data sets, no statistically significant finite-size variations are observed.  

\bibliography{Emulsion_paper_Bib}